\documentclass{aa}
\usepackage{graphicx}
\def\ssim{\setbox0=\hbox{$\propto$}%
\setbox1=\hbox{$<$}\dimen0=\ht1%
\advance\dimen0by-1.2pt\,\lower.6\dimen0%
\copy0\kern-\wd0\raise.4\dimen0\copy1 \,}

\def\gsim{\setbox0=\hbox{$\propto$}%
\setbox1=\hbox{$>$}\dimen0=\ht1%
\advance\dimen0by-1.2pt\,\lower.6\dimen0%
\copy0\kern-\wd0\raise.4\dimen0\copy1\,}

\def\lambdab{\lambda\mkern-9mu\lower1.2pt\hbox{$\mathchar'26$}}%

\begin{document}
   \title{Stellar evolution with rotation  VIII:}

\subtitle{Models at $Z$ = 10$^{-5}$ and CNO yields for
   early galactic evolution}

\author{Georges Meynet \& Andr\'e Maeder }

     \institute{Geneva Observatory CH--1290 Sauverny, Switzerland\\
              email: Georges.Meynet@obs.unige.ch \\
              email:  Andre.Maeder@obs.unige.ch      }

   \date{Received / Accepted }

\abstract{ We calculate a grid of star models with and without
the effects of axial rotation for stars in the mass range between 2 
and  60 M$_{\odot}$ for the metallicity $Z = 10^{-5}$.  
Star models with initial masses superior or equal to 9 M$_\odot$ were computed 
up to the end of the carbon--burning phase. Star models with masses between 2 and 7 M$_\odot$ were evolved beyond
the end of the He--burning phase through a few thermal pulses during the AGB phase.
Compared
to models at $Z=0.02$, the low $Z$ models
show faster rotating cores and stronger internal 
$\Omega$--gradients, which favour an important 
 mixing of the chemical elements. 
The enhancement of N/C at the surface may reach 2 to
3 orders of magnitude for fast rotating stars. Surface enrichments
may make the evolved stars less metal poor than they were initially.
In very low $Z$ models, primary nitrogen is produced
 during the He--burning phase
by rotational diffusion of $^{12}$C into the H--burning shell.
A large fraction of the primary $^{14}$N escapes further
destruction and enters the envelope of AGB stars, being ejected 
during the TP--AGB phase and the formation of a planetary nebula.
The intermediate mass stars of very low $Z$ are the main producers of 
primary $^{14}$N, but massive stars also contribute to this production;
no significant primary nitrogen is made in models at metallicity $Z$=0.004 
or above.
We calculate the chemical yields in He, C, N, O and heavy elements
and discuss the chemical evolution of the CNO elements 
at very low Z. Remarkably,
the  C/O vs O/H diagram is mainly sensitive to the
interval of stellar  masses, while the N/O vs O/H diagram
is mainly sensitive
to the average rotation of the stars contributing 
to the element synthesis.  The presently available
observations in these diagrams seem to  favour contributions  either from stars
down to about 2 M$_{\odot}$ with normal rotation velocities
or from stars above 8 M$_{\odot}$ but with very fast rotation.
\keywords  Physical data and processes: nucleosynthesis -- Stars: interiors, evolution, rotation -- Stars: early--types, AGB}

   \maketitle
%

\section{Introduction}

\begin{figure*}[tb]
  \resizebox{\hsize}{!}{\includegraphics{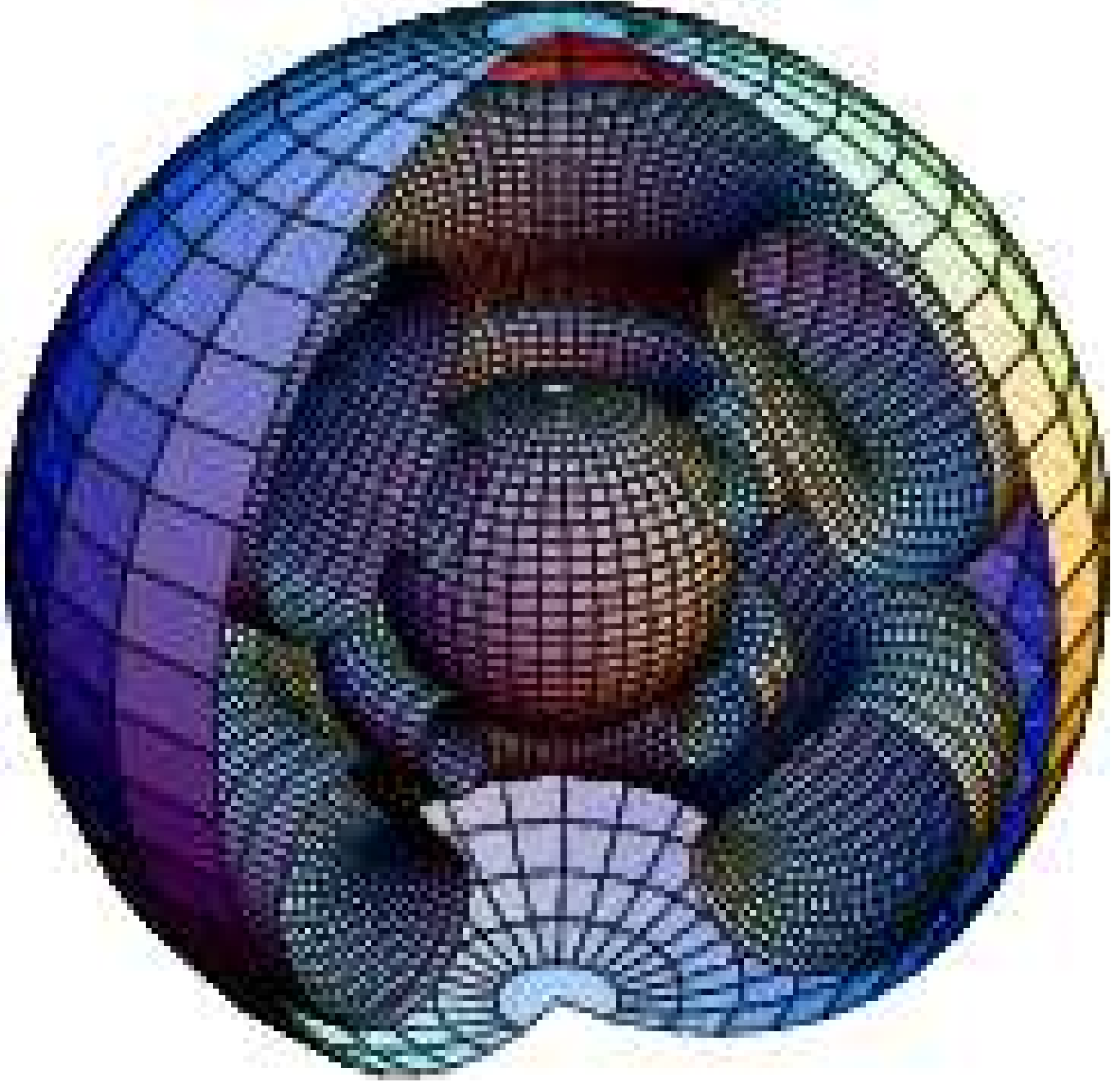}}
  \caption{Stream lines of meridional circulation in a rotating 20 M$_\odot$
model with  solar metallicity and
$v_{\rm ini}=300$ km s$^{-1}$ at the beginning of the H--burning phase (see text). 
The streamlines are in the meridian plane.
In the upper hemisphere on the right section, matter
is turning counterclockwise along the outer stream line and
clockwise along the inner one. The outer sphere is the star surface and has
a radius equal to 5.2 R$_\odot$. The inner sphere is the outer boundary of the
convective core. It has a radius of 1.7 R$_\odot$.}
  \label{circu}
\end{figure*}

Stellar rotation really modifies 
 all the outputs of stellar evolution for massive stars (Heger et al. \cite{heal00}; Heger \& Langer \cite{he00};
 Meynet \& Maeder \cite{MMV}). This is true at solar metallicity. At lower
metallicities, like  $Z$=0.004
in the SMC, we noticed that the effects of rotation
 are expected to be larger  
(Maeder \& Meynet \cite{MMVII}). In particular, for similar initial 
distributions of the rotational velocities, a larger fraction of the stars
at lower $Z$ reach break--up velocities. This is a result of the smaller
losses of angular momentum by the stellar winds. 

In addition, it may be that the initial distribution of the rotational
velocities is not the same at lower $Z$. Indeed, it has been shown
(Maeder et al. \cite{mgm})
that the fraction of Be stars (i.e. stars close to break--up)  is
much higher in the SMC than in the Milky Way. However, we do not 
know whether this is just  a consequence 
of the smaller mass loss, as said above, or whether
 the initial distribution of rotation velocities is also different
as a result of processes of star formation at low $Z$.
Whatever the exact origin of the higher fraction
of stars close to break--up at lower $Z$,
 this shows the need of studies 
of star models with rotation at low $Z$.

We consider here star models with metallicity $Z=10^{-5}$,
which is low enough to correspond to the most extreme
metallicity observed in halo stars of the order of
[Fe/H]$\simeq$ -3 and which nevertheless avoid the particularities 
of $Z = 0$ models, which we may  consider in a future paper.
The main possible comparisons
with the observations will concern the chemical evolution
of the abundances of the CNO elements and of other  heavy elements 
in halo stars and very low $Z$ galaxies.  This is why 
we put  here a particular
emphasis on the chemical yields in CNO at very low [Fe/H].
This is a topical point in relation with the problem
of primary nitrogen (Edmunds and Pagel \cite{edm}; 
Barbuy \cite{bar}; Carbon et al. \cite{car}; 
Thuan et al. \cite{thu} ; Izotov \& Thuan \cite{izo99}; Henry et al. \cite{Ha00}).
Also, the  recent debate around the behavior of the [O/Fe] ratio
at very low $Z$ (cf. Israelian et al. \cite{israelian01};
 Melendez et al. \cite{melendez}) 
shows the need of a better understanding of the CNO
yields at very low metallicities.  

In Sect. 2, we discuss the model physics. In Sect. 3,
we examine the internal rotation and the surface velocities in 
Sect. 4. The models with zero rotation are briefly mentioned
in Sect. 5. The HR diagram and lifetimes are discussed in Sect. 6. 
The evolution of surface abundances are examined in Sect. 7.
In Sect. 8, we discuss the problem of the origin of primary nitrogen and
we show how rotation can solve it. 
The chemical yields in  He, CNO and heavy elements are discussed in Sect. 9. 

\section{Physics of the models}

\begin{table}
\caption{Initial abundances in mass fraction.} \label{tbl-0}
\begin{center}\scriptsize
\begin{tabular}{cc}
\hline
             &                              \\
 Element     &    Initial abundance         \\
             &                              \\
\hline
             &                              \\
   H         &        0.76996750            \\
 $^3$He      &        0.00002438            \\
 $^4$He      &        0.22999812            \\
 $^{12}$C    &         7.5500e-7            \\
 $^{13}$C    &         0.1000e-7            \\
 $^{14}$N    &         2.3358e-7            \\
 $^{15}$N    &         0.0092e-7            \\
 $^{16}$O    &         67.100e-7            \\
 $^{17}$O    &         0.0300e-7            \\
 $^{18}$O    &         0.1500e-7            \\
 $^{20}$Ne   &         7.8368e-7            \\
 $^{22}$Ne   &         0.6306e-7            \\
 $^{24}$Mg   &         3.2474e-7            \\
 $^{25}$Mg   &         0.4268e-7            \\
 $^{26}$Mg   &         0.4897e-7            \\
             &                              \\
 \hline
             &                              \\
\end{tabular}
\end{center}

\end{table}

The initial composition is given in Table~\ref{tbl-0}.
The composition 
is enhanced in $\alpha$--elements. 
As in paper VII, the opacities are
from Iglesias \& Rogers (\cite{IR96}), complemented at low
temperatures with the molecular opacities of Alexander
(http://web.physics.twsu.edu/alex/wwwdra.htm). The nuclear
 reaction rates are also the same as in paper VII and are
 based on the new NACRE data basis (Angulo et al. \cite{Ang99}).

The physics of the present models at $Z=10^{-5}$ is the same
as for models at $Z$=0.004 (Maeder \& Meynet \cite{MMVII}).
For rotation, the hydrostatic effects and the surface distortion
are included (Meynet \& Maeder \cite{MMI}), so that the $T_{\mathrm{eff}}$
given here corresponds to an average orientation angle. The
diffusion by shears,
which is the main effect for the mixing of chemical elements,
 is included (Maeder \cite{MII}), with 
account of the effects of the horizontal turbulence,
which reduces the shear effects in regions of steep 
$\mu$--gradients and reinforces it in regions of low 
$\mu$--gradients (Maeder \& Meynet \cite{MMVII}). 

Meridional circulation is the main effect for the 
internal transport of angular momentum. We use here the
expression by Maeder \& Zahn (\cite{MZIII}) for
the vertical component $U(r)$ of the meridional circulation.
It is interesting to represent graphically this circulation.
Fig.~\ref{circu} illustrates the patterns of the 
meridional circulation in a 20 M$_{\odot}$ star at $Z=0.020$
and initial rotation velocity $v_{\rm ini}=300$ km s$^{-1}$
on the ZAMS.
The figure is symmetrical with respect
to the rotation axis, as well with respect to the equatorial plane.
The small inner sphere is the edge of the central convective core.
The inner tube, in the upper hemisphere, 
represents an interior cell of the meridional circulation.
There is an ensemble of such concentric
 tubes   with different meridional velocities.
The motions occur in a meridian plane (i.e. turning around 
the tube). 
In the upper hemisphere and around the inner tube,
the fluid elements go upward
 on the inner side of the tube and descend toward the equator on the 
outer side of the tube ($U(r)$ is positive). 
The external tube represents an outer circulation cell,
due to the  Gratton--\H{O}pik term, which is important in the outer stellar
layers. 
This term leads to a negative  $U(r)$, which means
that, in the upper hemisphere, the fluid goes up on the outer side
of the tube and down along the inner side.
There also, this tube is one among an ensemble of stream lines turning
in the meridian plane.

The mass loss rates are based on the same references as in the paper for
the $Z$=0.004 models (Maeder \& Meynet \cite{MMVII}) and in particular
on the data by Kudritzki \& Puls (\cite{KP00}) for the OB stars.
 Of course, the
strong reduction of the mass loss rates with
metallicity for stars below 60 M$_{\odot}$ makes the mass
loss rather unimportant for the metallicity $Z=10^{-5}$
 considered here, as illustrated by Table 1
which shows the values of the final masses.
 We account for 
the effects of rotation on the mass loss rates, according to
 the standard stellar wind theory applied to a rotating star
(Maeder \& Meynet \cite{MMVI}). The net result is
that the very massive stars with initial M $\geq$ 60 M$_{\odot}$
may still experience significant mass loss as shown in Table 1,
 if they rotate very fast. 

Star models with initial masses superior or equal to 9 M$_\odot$ were computed 
up to the end of the carbon--burning phase. Star models with masses between 2 and 7 M$_\odot$ were evolved beyond
the end of the He--burning phase through a few thermal pulses during the AGB phase.

\begin{figure}[tb]
  \resizebox{\hsize}{!}{\includegraphics{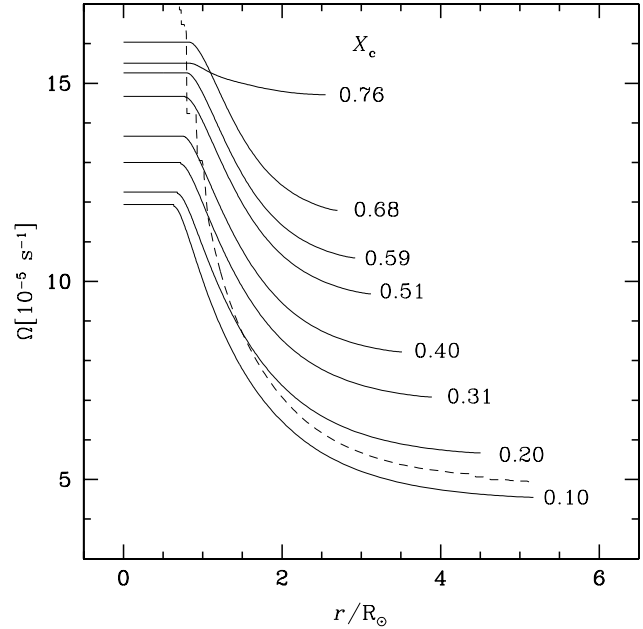}}
  \caption{Evolution of the angular velocity $\Omega$ 
 as a function of the distance to the center
in a 15 M$_\odot$ star with $v_{\rm ini}$ = 300 km s$^{-1}$ and
$Z = 10^{-5}$. 
$X_{\rm c}$ is the hydrogen mass fraction at the center.
The dotted line shows the profile when the He--core contracts at the end
of the H--burning phase.}
  \label{omegar}
\end{figure}

\begin{figure}[tb]
  \resizebox{\hsize}{!}{\includegraphics{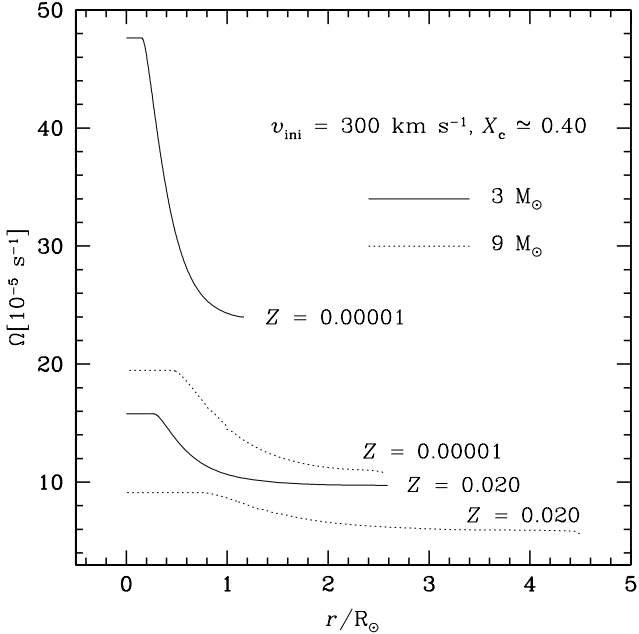}}
  \caption{Variation of the angular velocity $\Omega$ 
 as a function of the distance to the center
in 3 and 9 M$_\odot$ star models with $v_{\rm ini}$ = 300 km s$^{-1}$ at $Z =
0.020$ and
$Z = 10^{-5}$. The mass fraction of hydrogen at the centre $X_{\rm c} \simeq
0.40$.
}
  \label{omegac}
\end{figure}

\section{The evolution of the internal rotation
and meridional circulation}

There are remarkable differences in the internal distributions
of the angular velocity $\Omega(r)$ depending on the stellar metallicity $Z$.
This was already suggested in paper VII (Maeder and Meynet
\cite{MMVII}), when comparing models at $Z = 0.004$
and $Z = 0.020$. It is  extended here with models
at $Z = 10^{-5}$. 

These matters   are not academic problems !  Indeed, the distribution
of $\Omega(r)$ determines for example the mixing of chemical elements,
the size of the convective core and therefore the chemical yields.
The results in Sect. 8 and 9 below on the chemical yields
are a consequence of the distribution of $\Omega(r)$.

Fig.~\ref{omegar} shows the evolution of $\Omega(r)$ during the
MS phase of a 15 M$_{\odot}$ at  $Z = 10^{-5}$, (this follows the initial
convergence of the $\Omega$--profile which is very short, i.e. $\leq$ 1\%
of the MS lifetime).  We notice that the rotation of the
 convective core only has a small decrease during the MS phase, much
smaller than at higher metallicities. This results from 2 effects.
a) The mass loss at  $Z = 10^{-5}$ is much smaller than at solar composition
and thus less angular momentum is removed from the star. b) As we shall
see below, the meridional circulation  is very slow in the outer regions 
of the models at very low $Z$
and it transports much less angular momentum outwards than in models
at solar composition. In view of these remarks,  it is 
likely that massive stars at lower $Z$ have faster spinning cores.

\begin{figure}[tb]
  \resizebox{\hsize}{!}{\includegraphics{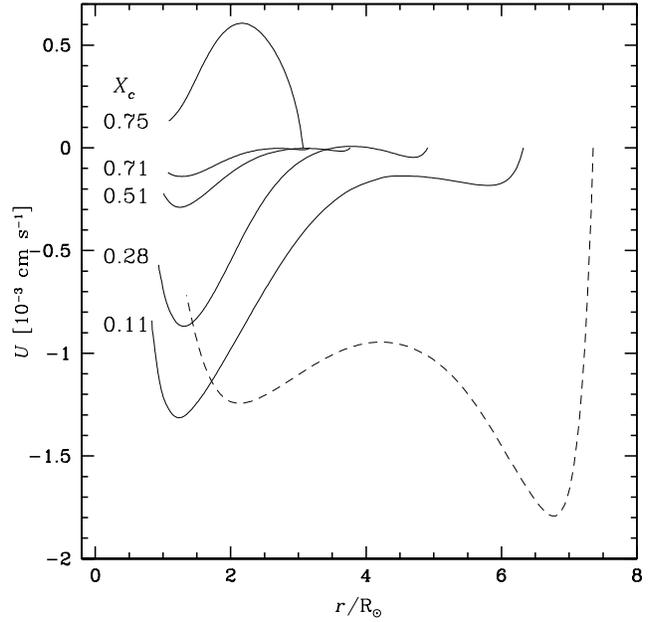}}
  \caption{Evolution of $U(r)$ the radial term of the vertical 
component of the
velocity of meridional circulation for a model of 20 M$_{\odot}$
with $Z = 10^{-5}$ at various stages during the MS phase.
 $X_c$ is the hydrogen mass 
fraction at the center. The dashed line shows the values of $U(r)$ inside
a 20 M$_\odot$ model at $Z$ = 0.004 when $X_{\rm c}$ = 0.28.}
  \label{Ur}
\end{figure}

Another significant difference shown by Fig.~\ref{omegar}
concerns  the  gradient of $\Omega$ outside the convective cores.
Here, the gradients are steeper and they remain significant 
up to the stellar surface, while at $Z = 0.02$ the $\Omega$--distribution
becomes very flat in the external layers, as evolution proceeds
(Meynet \& Maeder \cite{MMV}).
This difference is well illustrated in Fig.~\ref{omegac}, where 
we  notice for the 3 and 9 M$_{\odot}$ models the
much steeper $\Omega$--gradients at lower $Z$, while the models at $Z=0.02$ 
show very flat gradients in the outer layers.
The reason for the higher $\Omega$--gradients here are the same 
as for the faster spinning cores. These  higher $\Omega$--gradients
imply stronger shears and thus more mixing by shear diffusion,
which is the main effect for the outward transport of the chemical species.
(The differences in $\Omega$ between the 9 and 3 M$_{\odot}$
models result from the fact that we consider stars with the
same $v_{\rm ini}$, but different radii).

Fig.~\ref{Ur} shows an example  at $Z = 10^{-5}$ of 
the evolution of $U(r)$, the vertical component
of the velocity of meridional circulation. The size and evolution 
of $U(r)$ is very different from the case at $Z = 0.02$. At $Z = 0.02$,
$U(r)$ takes large negative values particularly
in the outer layers.  This is due to their low density, which makes
a large  Gratton--\H{O}pik term 
  $\frac{-\Omega^2}{2 \pi G \overline{\rho}}$
in the expression of $U(r)$, (cf. Maeder and Zahn \cite{MZIII}).
At $Z = 10^{-5}$, the large negative values of $U(r)$ have disappeared,
$U(r)$ is equal to $10^{-3}$ cm s$^{-1}$ at the end of the MS
phase, while it was 50 times more negative in the corresponding models at 
$Z = 0.02$ (Meynet \& Maeder \cite{MMV}). The differences do not concern so much the deep interior,
but mainly the outer layers. The physical reason of the above
differences is the fact that the star is more compact at lower
$Z$ and that  the density in the outer layers is not as low
as at solar composition.

Fig.~\ref{Ur} also shows the curious curve for  a model at
$Z = 0.004$. In the interior, $U(r)$ is about the same as 
in the present models (and this is true for all $Z$ values).
The big external dip of $U(r)$, which was present at
$Z = 0.02$ is very much reduced, but still present at $Z = 0.004$,
while at $Z = 10^{-5}$  the external dip is fully absent.

The small $U(r)$ in the external layers of the present models
is mainly responsible for the presence of an $\Omega$--gradient 
up to the stellar surface (cf. Fig.~\ref{omegar}).
Since the mixing of the chemical elements is mainly driven by the shear,
the presence of this $\Omega$--gradient in the outer layers enables
the large mixing and surface chemical enrichments
that  are present in the $Z = 10^{-5}$ models.

\section{The evolution of the surface rotation velocities}

\begin{figure}[tb]
  \resizebox{\hsize}{!}{\includegraphics{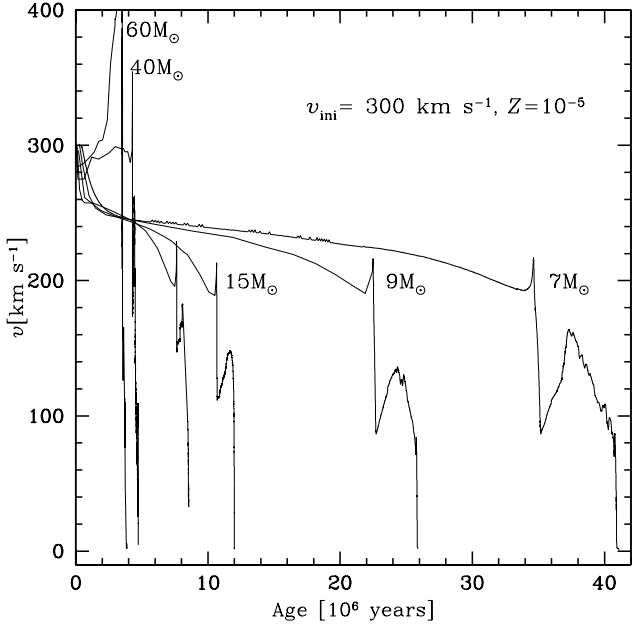}}
  \caption{Evolution of the surface equatorial velocity as 
a function of time for stars of different initial masses
with $v_{\mathrm {ini}}$ = 300 km s$^{-1}$ and $Z =10^{-5}$.
The track without label corresponds to a 20 M$_\odot$ model. 
}
  \label{vage}
\end{figure}

\begin{figure}[tb]
  \resizebox{\hsize}{!}{\includegraphics{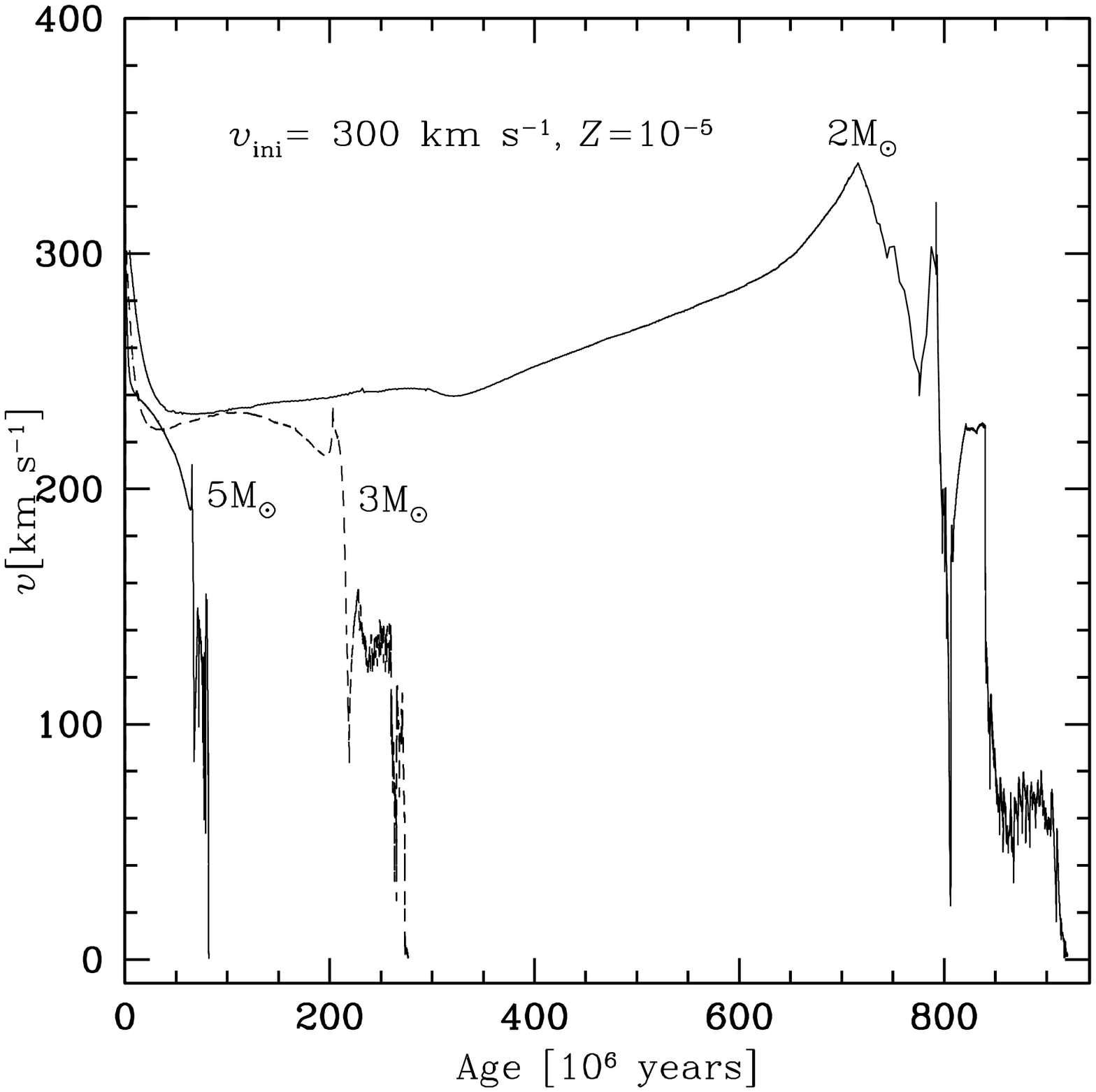}}
  \caption{Same as Fig.~\ref{vage} for a 2, 3 and 5 M$_\odot$ 
stellar model.}
  \label{vage2}
\end{figure}

\begin{figure}[tb]
  \resizebox{\hsize}{!}{\includegraphics{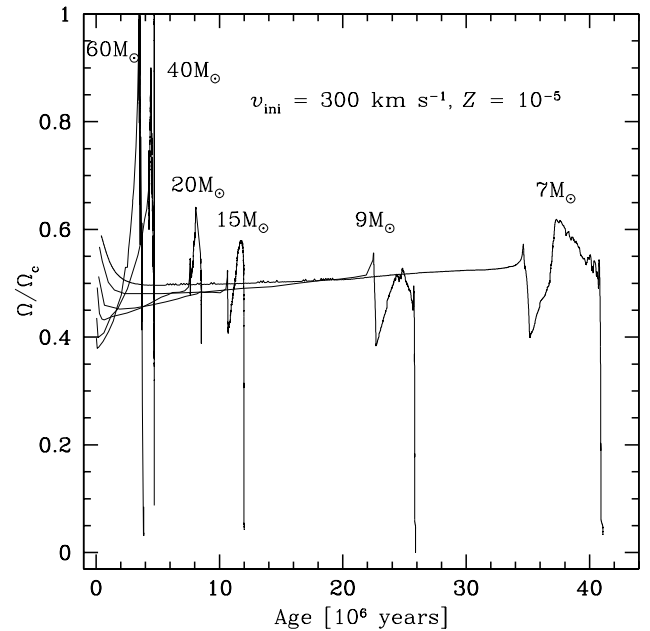}}
  \caption{Evolution of the ratio $\Omega/\Omega_{\rm c}$ 
of the angular velocity to the break--up angular velocity
at the stellar surface for stars of different masses 
at $Z = 10^{-5}$.
}
  \label{omegage}
\end{figure}

\begin{figure}[tb]
  \resizebox{\hsize}{!}{\includegraphics{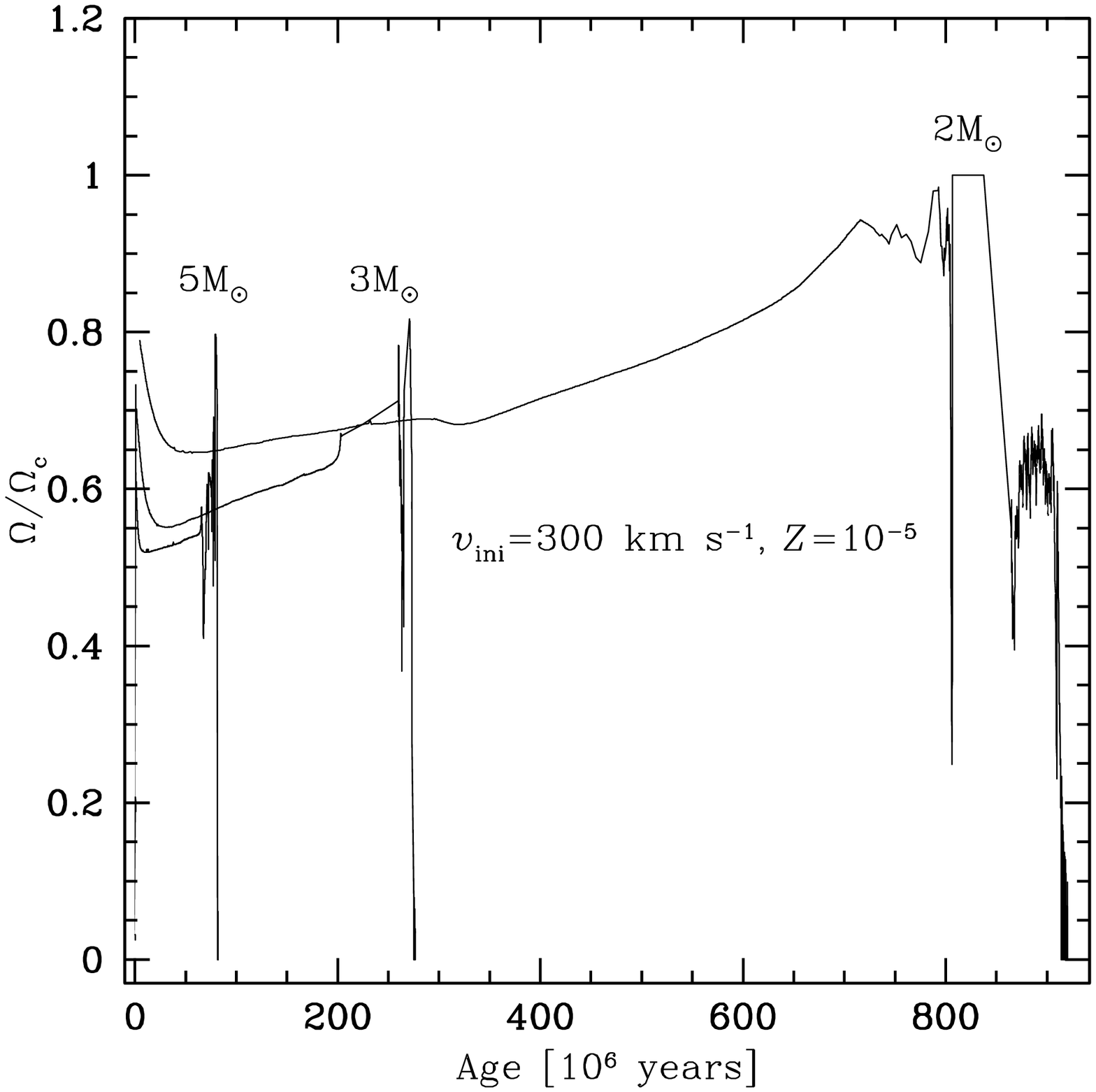}}
  \caption{Same as Fig.~\ref{omegage} for a 2, 3 and 5 M$_\odot$ 
stellar model.
}
  \label{omegag2}
\end{figure}

\begin{figure}[tb]
  \resizebox{\hsize}{!}{\includegraphics{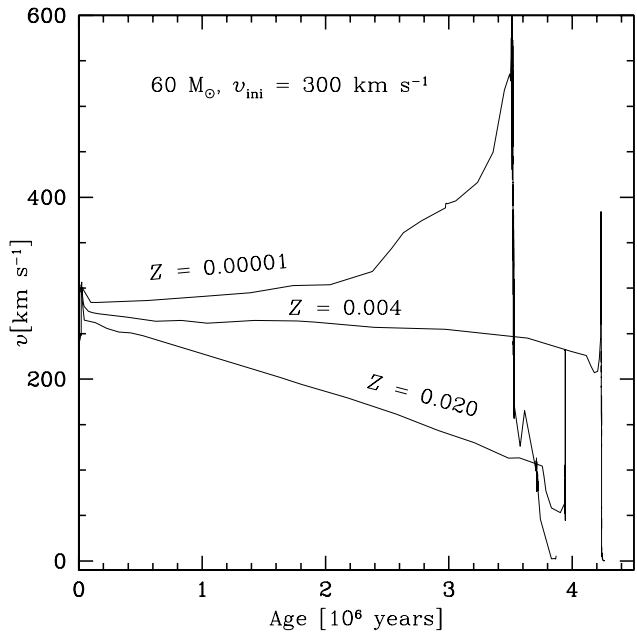}}
  \caption{Evolution of the surface equatorial velocity as 
a function of time for 60 M$_\odot$ stars with $v_{\rm ini}$ = 300
km s$^{-1}$ at different initial metallicities.
}
  \label{v60}
\end{figure}



It will probably be a certain time until we are able to observe $v \sin i$
for  stars in galaxies with $Z = 10^{-5}$. Nevertheless, these objects have
contributed to shape the composition of our universe and they
deserve some interest.

Fig.~\ref{vage} and Fig.~\ref{vage2} show the evolution
of the surface rotational velocities for models
 with initial masses from 2 to 60 M$_{\odot}$. 
Figs.~\ref{omegage} and   \ref{omegag2} show
the corresponding evolution of  $\frac{\Omega}{\Omega_{\rm c}}$.
We notice the relative 
 constancy of $\frac{\Omega}{\Omega_{\rm c}}$ during the
MS evolution for  stars with mass between 5 and  20 M$_{\odot}$.
The cases of 40 and 60 M$_{\odot}$ are noticeable as shown by 
Figs.~\ref{vage} and \ref{omegage}. These models reach
 the break--up velocities near the end of the MS phase. This
is completely different  from the models at $Z = 0.02$, where 
the rotation becomes very small due to the huge losses of
mass and angular momentum. Thus, if the initial mass function
at low $Z$ extends up to high mass stars, as often supposed,
rotation is likely to be a major effect in the course of the
evolution of massive stars,
 since many of them are likely to reach break--up velocities.
This would even more be the case for the massive stars which have
a blueward evolution as a result
of strong internal mixing. Their radii would decrease, thus favouring
extreme rotation velocities.
We note that for masses 
between 3 and 20 M$_{\odot}$, the rotation velocity keeps about constant 
during the MS phase, before decreasing in the post--MS phases.

Fig.~ \ref{v60} clearly illustrates 
the very different evolution of the rotational velocities
of a 60 M$_{\odot}$ at various metallicities.
At low $Z$ like in the  models at $Z = 10^{-5}$ ,
 the growth of $\frac{\Omega}{\Omega_{\rm c}}$
is possible because of the very small mass loss and
also it is favoured
by the outward transport of
angular momentum which is much larger for the more massive stars.
As shown by Maeder and Meynet (\cite{MMVII}), the values of 
$U(r)$ are more negative for the larger stellar  masses, 
due to several facts: lower gravity, higher radiation pressure, larger
L/M ratio and especially the lower density. Thus, the outward transport
is more efficient.

In view of these results, we may somehow precise our suggestion
 (Maeder and Meynet \cite{MMVII}) that at very low $Z$  a large fraction
of the massive stars reach their break--up velocities.  This seems
true for the highest masses above about  
30 M$_{\odot}$, but not necessarily
for the OB stars below this limit. This question is of high
importance, because if the massive stars reach their break--up velocity,
most of their evolutionary and structural  properties will be affected.
For example, they could also
 lose a lot of mass and produce some Wolf--Rayet
(WR) stars. They would have a relatively small remaining mass
at the time of the supernova explosion, like their counterparts at
solar composition.

For the models of 3 M$_{\odot}$, as illustrated in 
Figs.~\ref{vage2} and \ref{omegag2}, the rotation velocity 
remains about constant during the
MS phase, while $\frac{\Omega}{\Omega_c}$ increases.
For 2  M$_{\odot}$, we notice a net increase.
This particular behavior is due to the different shape
 of the track of the 2 M$_{\odot}$ model
 in the HR diagram (Fig.~\ref{HR}), which mimics
 the tracks of lower masses dominated by the pp chain. 
This is well explainable, because
at $Z = 10^{-5}$ the CNO cycle is less important than at
solar composition, thus the mass limit where the CNO cycle 
starts dominating over the pp chain is shifted upward.
For this model of 2 M$_{\odot}$, there is no large increase 
 of the radius during the MS evolution  and thus rotation keeps higher.

When one examines the evolution of $v \sin i$ during the MS
phase for stars of the same mass but 
 different initial velocities, one usually notes  at 
Z= 0.02 a so--called velocity convergence (Langer \cite{Lan98}).
This is due to the fact that the faster
rotating stars lose more mass and thus more angular momentum.
In the present models, the mass loss rates are in general
very small (as long as the stars are not at break--up)
and thus there is no velocity convergence, i.e. the stars of
different initial velocities finish the MS phase with different
velocities as illustrated by Table 1.

\section{Models with zero rotation}

\begin{figure*}[tb]
  \resizebox{\hsize}{!}{\includegraphics[angle=-90]{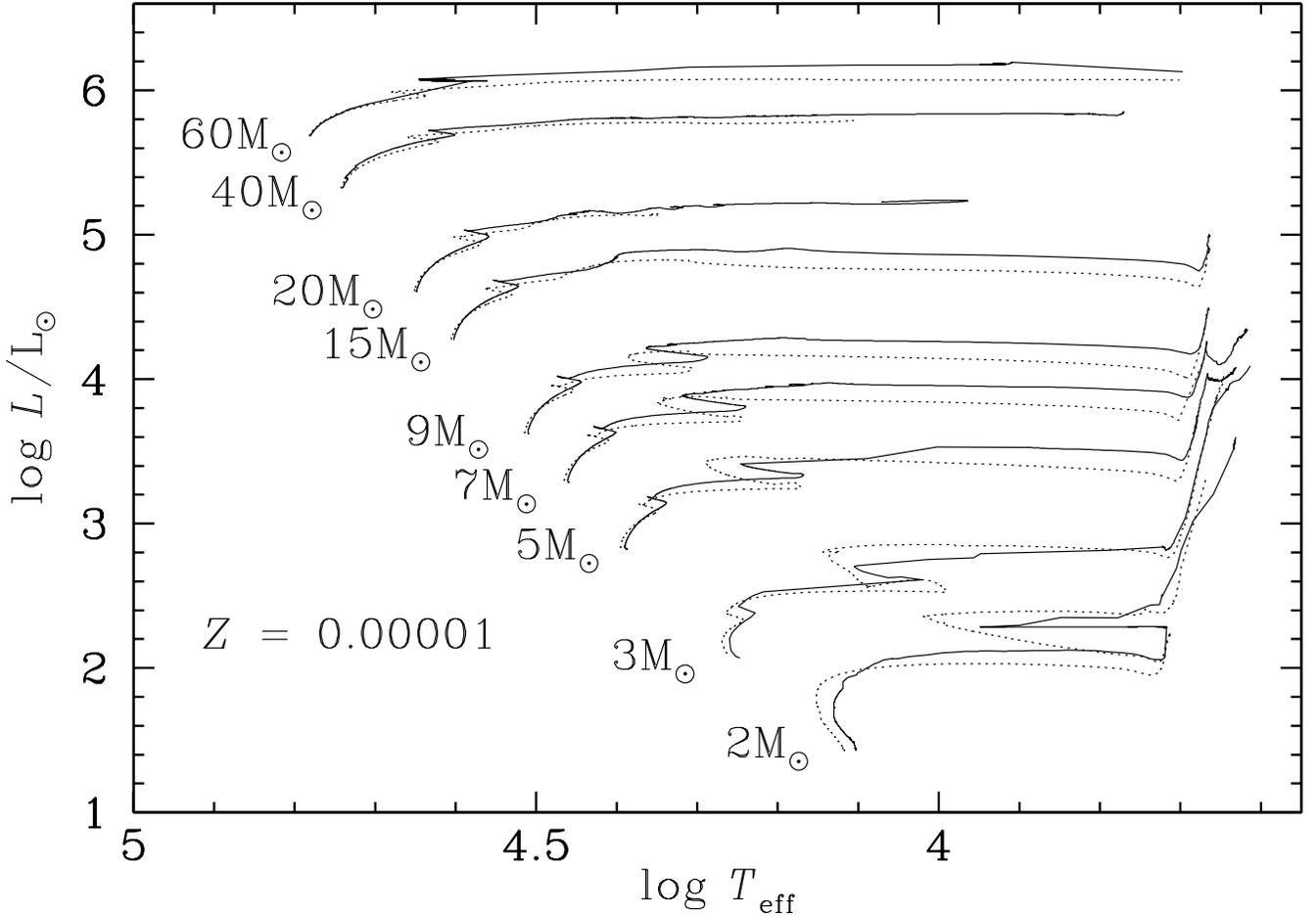}}
  \caption{Evolutionary tracks for non--rotating 
(dotted lines) and rotating (continuous lines) models for a metallicity $Z =
10^{-5}$. 
The rotating models
have an initial velocity $v_{\rm ini}$ of 300 km s$^{-1}$.}
  \label{HR}
\end{figure*}

For purpose of comparison, we have computed non--rotating stellar models
with the same physical ingredients as for the computation of the
rotating ones. The evolutionary tracks are presented in Fig.~\ref{HR},
the lifetimes and various properties of the models are given in
Table~\ref{tbl-1}.
The models were computed with the Schwarzschild criterion for convection and
therefore present the usual differences when compared with models accounting
for the effect of overshooting (see paper V and VII for a more detailed
discussion).

With respect to our previous grids of stellar models 
at solar metallicity (see paper V), the present models are shifted toward 
higher effective temperatures, by about 0.15--0.20 dex in log $T_{\rm eff}$. 
The stars are much more compact than at solar metallicity by nearly a factor
two
(more precisely by a factor between 1.8--1.9 depending on the initial mass). 
This is a well known consequence of the low opacities in the outer layers of
metal poor stars.
 
In paper VII, we 
noticed that models at $Z$ = 0.004 with initial masses between 10 and 12.5 M$_\odot$ were
showing a behavior in the HR diagram
intermediate between the cases of stars presenting
 a well developed blue loop and 
the case of more massive stars,
 which do not produce any blue loop,
 but begin to burn their helium in their core
at a high effective temperature, while
 they cross the HR diagram for the first
 time. Here at $Z = 10^{-5}$, this transition zone covers a broader range of initial masses, from
2 to about 15 M$_\odot$. This is well consistent with grids
of stellar models computed by other authors. Indeed a similar broadening in
mass
for this transition region can be observed in the grids by the Padova group
(compare for instance the grids by Fagotto et al. \cite{fa94} at $Z$ = 0.0004 and that
of Girardi et al. \cite{gi96} at $Z$ = 0.0001).

From Fig.~\ref{HR}, one sees that the 20 M$_\odot$ model does not reach the red
supergiant 
phase at least before the end of the C--burning phase. This feature is also
present
in zero metallicity stellar models (see e.g. Marigo et al. \cite{Ma01}), but for
pop III models, it
extends over
a broader range of initial masses (from $\sim$ 10 to $\sim$ 40 M$_\odot$).
Finally, let us note that,
when the metallicity decreases, the mass limit for the helium flash decreases
(see also Marigo et al. \cite{Ma01}).
This is a consequence of the higher central temperatures reached in metal poor
stars.
Typically a 2 M$_\odot$ model at solar metallicity,  computed with the same physical
ingredients as 
used in the present work, undergoes He--flash, while the corresponding model at
$Z =10^{-5}$
ignites helium in a non--degenerate environment.

\section{HR diagram, mass--luminosity relations and lifetimes}

\begin{table*}
\caption{Properties of the stellar models at the end of 
the H--burning phase, at the end of the He--burning phase and at the end of the C--burning phase or during the thermal pulse--AGB phase. 
The masses are in solar mass, the velocities, in km s$^{-1}$ and the lifetimes, in
million years. The abundances are in mass fraction.
The abundance ratios 
are normalized to their initial values, which are, in mass fraction, (N/C)$_{\rm ini}$= 0.309 and (N/O)$_{\rm ini}$= 0.035 .} \label{tbl-1}
\begin{center}\scriptsize
\begin{tabular}{ccc|ccccc|ccccc|ccccc}
\hline
    &     &     &       &       &      &      &      &        &     &      &    &       &        &        &     &      &              \\
    &  & & \multicolumn{5}{|c|}{End of H--burning}&\multicolumn{5}{|c|}{End of He--burning} &\multicolumn{5}{|c} {End of C--burning} \\
    &     &     &       &       &      &      &      &        &     &      &      &       &        &        &     &      &             \\
$M$ & $v_{\rm ini}$ &  $\overline{v}$  & $t_{\mathrm{H}}$ & $v$ & $Y_{\mathrm{s}}$ & N/C & N/O & $t_{\mathrm{He}}$ & $v$ & $Y_{\mathrm{s}}$ 
& N/C   & N/O & $M_{\rm fin}$ & $v$ &
  $Y_{\mathrm{s}}$ & N/C   & N/O    \\
    &     &     &       &       &      &      &      &        &     &      &      &       &       &     &      &      &        \\
\hline
    &     &     &       &       &      &      &      &        &     &      &      &       &       &     &      &      &        \\
 60 &  0  & 0   & 3.883 &   0   & 0.23 & 1.00 & 1.00 & 0.332  &  0  & 0.23 & 1.00 &  1.00 & 59.57 & 0   & 0.23 & 1.00 & 1.00   \\
    &300  & 327 & 3.513 & 545   & 0.35 & 64.2 & 25.8 & 0.345  & 2   & 0.47 & 126  &  49.8 & 50.42 & 5   & 0.51 & 145  & 63.0   \\
    &     &     &       &       &      &      &      &        &     &      &      &       &       &     &      &      &        \\
 40 &  0  & 0   & 4.974 &   0   & 0.23 & 1.00 & 1.00 & 0.419  &  0  & 0.23 & 1.00 &  1.00 & 39.86 & 0   & 0.23 & 1.00 & 1.00   \\
    &300  & 289 & 4.279 & 355   & 0.25 & 80.8 & 14.3 & 0.436  & 21  & 0.28 & 123  &  19.1 & 38.61 & 4   & 0.30 & 139  & 22.4   \\
    &     &     &       &       &      &      &      &        &     &      &      &       &       &     &      &      &        \\
 20 &  0  & 0   & 8.773 &   0   & 0.23 & 1.00 & 1.00 & 0.886  &  0  & 0.23 & 1.00 &  1.00 & 19.97 & 0   & 0.23 & 1.00 & 1.00   \\
    &200  & 157 & 7.445 & 139   & 0.23 & 12.8 & 4.89 & 0.978  & 8   & 0.23 & 21.3 &  6.55 & 19.97 & 1   & 0.26 & 45.2 & 11.6   \\
    &300  & 240 & 7.624 & 228   & 0.23 & 28.6 & 7.20 & 0.902  & 81  & 0.24 &54.2  &  10.4 & 19.97 & 75  & 0.24 & 54.8 & 10.6   \\
    &400  & 325 & 7.737 & 338   & 0.23 & 69.1 & 9.23 & 0.932  & 155 & 0.27 & 150  &  16.6 & 19.90 & 208 & 0.27 & 153  & 17.3   \\
    &     &     &       &       &      &      &      &        &     &      &      &       &       &     &      &      &        \\
 15 &  0  & 0   & 12.12 &   0   & 0.23 & 1.00 & 1.00 & 1.473  &  0  & 0.23 & 1.00 &  1.00 & 14.99 & 0   & 0.30 & 42.3 & 19.5   \\
    &300  & 234 & 10.65 & 212   & 0.23 & 21.9 & 6.74 & 1.281  & 111 & 0.24 & 53.6 &  11.3 & 14.98 & 2   & 0.29 & 126  & 23.2   \\
    &     &     &       &       &      &      &      &        &     &      &      &       &       &     &      &      &        \\
  9 &  0  & 0   & 25.12 &   0   & 0.23 & 1.00 & 1.00 & 3.285  &  0  & 0.23 & 1.00 &  1.00 & 8.998$^a$ & 0   & 0.24 & 34.5 & 15.5   \\
    &200  & 151 & 22.03 & 128   & 0.23 & 5.45 & 3.16 & 3.414  & 72  & 0.23 & 24.6 &  8.00 & 8.997 & 1   & 0.26 & 79.3 & 18.0   \\
    &300  & 230 & 22.50 & 212   & 0.23 & 79.5 & 10.5 & 3.040  & 73  & 0.23 & 107  &  12.0 & 8.997$^a$ & 2   & 0.27 & 234  & 23.4   \\
    &400  & 307 & 22.77 & 294   & 0.23 & 365  & 13.7 & 3.612  & 86  & 0.24 & 477  &  17.4 & 8.997$^a$ & 2   & 0.27 & 579  & 25.7   \\
    &     &     &       &       &      &      &      &        &     &      &      &       &       &     &      &      &        \\
    &  & & \multicolumn{5}{|c|}{ }&\multicolumn{5}{|c|}{ } &\multicolumn{5}{|c}{AGB phase} \\
    &     &     &       &       &      &      &      &        &     &      &      &       &       &     &      &      &        \\
    &     &     &       &       &      &      &      &        &     &      &      &       &  $M_{\rm fin}$  & $Y_{\mathrm{s}}$  &   C   &  N & O       \\
    &     &     &       &       &      &      &      &        &     &      &      &       &       &     &      &      &         \\
  7 &  0  & 0   & 38.28 &   0   & 0.23 & 1.00 & 1.00 & 5.917  &  0  & 0.23 & 1.00 &  1.00 & 6.999 & 0.24 &  2.5E-7 & 2.4E-6 & 4.9E-6 \\
    &300  & 229 & 34.66 & 209   & 0.23 & 78.6 & 9.67 & 5.576  & 74  & 0.24 & 160  &  12.9 & 6.998 & 0.36 &  1.7E-3 & 7.0E-4 & 6.6E-4 \\
    &     &     &       &       &      &      &      &        &     &      &      &       &       &      &         &        &        \\
  5 &  0  & 0   & 70.71 &   0   & 0.23 & 1.00 & 1.00 & 14.38  &  0  & 0.23 & 1.00 &  1.00 & 5.000 & 0.24 &  2.5E-7 & 2.2E-6 & 5.0E-6 \\
    &300  & 226 & 65.69 & 198   & 0.23 & 16.3 & 5.32 & 13.61  & 94  & 0.24 & 68.2 &  12.7 & 4.996 & 0.34 &  7.5E-4 & 1.0E-3 & 3.9E-4 \\
    &     &     &       &       &      &      &      &        &     &      &      &       &       &      &         &        &        \\
  3 &  0  & 0   & 200.6 &   0   & 0.23 & 1.00 & 1.00 & 50.22  &  0  & 0.23 & 1.00 &  1.00 & 2.990 & 0.27 &  2.9E-7 & 1.7E-6 & 5.5E-6 \\
    &300  & 229 & 208.0 & 228   & 0.23 & 26.9 & 6.97 & 52.52  & 29  & 0.26 & 177  &  15.8 & 2.910 & 0.29 &  1.3E-5 & 7.4E-4 & 2.0E-4 \\
    &     &     &       &       &      &      &      &        &     &      &      &       &       &      &         &        &        \\
  2 &  0  & 0   & 637.5 &   0   & 0.23 & 1.00 & 1.00 & 109.3  &  0  & 0.24 & 4.47 &  2.59 & 2.000 & 0.25 &  3.5E-7 & 1.0E-6 & 6.3E-6 \\
    &300  & 251 & 688.3 & 318   & 0.23 & 3.27 & 2.29 & 107.8  &  9  & 0.28 & 257  &  10.8 & 1.863 & 0.29 &  6.2E-8 & 6.0E-6 & 5.7E-6 \\
    &     &     &       &       &      &      &      &        &     &      &      &       &       &      &         &        &        \\
\hline
 \multicolumn{18}{l} {    } \\
 \multicolumn{18}{l}{$^a$ Models at the beginning of the C--burning phase.}
\end{tabular}
\end{center}

\end{table*}

The effects of rotation at $Z=0.020$
have already been discussed in 
Talon et al. (\cite{Tal97}), Denissenkov et al. (\cite{de99}), Heger et al. (\cite{heal00}), Heger \& Langer (\cite{he00}),
Meynet \& Maeder (\cite{MMV}). At the metallicity of the Small Magellanic Cloud, the effects
of rotation have been discussed by Maeder \& Meynet (\cite{MMVII}).
Let us very briefly recall the most important effects of rotation:
\begin{itemize}

\item The Main Sequence (MS) evolutionary tracks with rotation are extended
toward lower effective temperatures and reach higher luminosities at the
end of the MS phase than their
non--rotating counterparts.
In that respect rotation acts as a moderate overshoot.

\item The MS lifetimes increase with rotation, typically by about 10\% for
a 20M$_\odot$ model with an average rotational velocity
on the MS corresponding to the observed ones. 

\item For a given value of the initial mass and metallicity, the evolutionary
tracks may be different, due
to different initial rotational velocities. Even for a given initial rotation,
the tracks may appear
different depending on the angle of view, since the polar regions of a
rotating star are in general
hotter than the equatorial ones (cf. Maeder \& Peytremann \cite{ma70}). This induces some scatter in the position
of the end of the MS phase in the HR diagram.

\item The theoretical period--luminosity relation for the Cepheids is changed by rotation.
A Cepheid, at a given position in the
HR diagram, if originating from a rotating progenitor, will pulsate with a
longer period than 
a Cepheid having a
non--rotating progenitor.

\item The evolution toward the red supergiant phase is favoured by rotation. Rotation also
facilitates the formation of Wolf--Rayet stars.

\item The surface abundances are modified.

\end{itemize}
We shall see that the models 
at the very low metallicity $Z = 10^{-5}$
present some striking differences with respect
to what happens at higher metallicities:
in particular, 
rotation implies smaller main sequence lifetimes
 and rotation does not favour the evolution toward the red
supergiant stage (at least for the range of initial
velocities explored here).

\subsection{The HR diagram}

For most of the stellar models, we
computed the rotating tracks for an initial velocity $v_{\rm ini}$ = 300 km
s$^{-1}$. This value of
 $v_{\rm ini}$ corresponds to a mean
velocity  $\overline {v}$ during the MS
 between 226 and 240 km s$^{-1}$ for
initial masses below 20 M$_\odot$ (see Table~\ref{tbl-1}).
These values are close to the mean rotational velocities observed for OBV
type stars at solar metallicity, which are between 200--250 km s$^{-1}$. 
For more massive stellar models at $Z= 10^{-5}$
the average velocities are higher. This results essentially from the larger
outward transport of angular momentum by circulation in more massive stars
(see Section 4 and Maeder \& Meynet \cite{MMVII}).

Figure~\ref{HR} shows the evolutionary tracks of non--rotating and rotating
stellar models for initial masses between 2 and 60 M$_\odot$.
The effective temperatures plotted correspond
to an average orientation angle (see also paper I). 
At the beginning of the evolution  on the ZAMS, 
rotational mixing has no impact on the
structure, since the star is homogeneous. At this stage,
only the hydrostatic effects
of rotation are present, {\it i.e.} the
effects due to the centrifugal acceleration term in the 
hydrostatic equilibrium equation. As is well known, these effects shift the ZAMS
position toward
lower values of $L$ and $T_{\rm eff}$ (see e.g. paper I). From
Fig.~\ref{HR}, one sees that the less massive the star, the greater
the shift. This
results
from the facts that, for a given $v_{\rm ini}$, the lower the initial mass,
the greater the ratio of the centrifugal force to the gravity. Indeed
this ratio, equal to ${v_{\rm ini}^2\over R} { R^2 \over GM}$, 
varies as about  $1/M^{\alpha}$ with $\alpha$ equal to about 0.4.

As was the case at higher metallicities,  
the MS width is increased by rotation. 
Rotational mixing
brings fresh H--fuel into the convective
core, 
slowing down its decrease in mass
during the MS. A more massive He--core
is produced at the end of the H--burning phase, which
favours the extension of the tracks toward lower effective temperatures. 
Rotational mixing also transports
helium and other H--burning products (essentially nitrogen)
into the radiative envelope. The He--enrichment lowers the opacity. This
contributes to the more rapid increase of the stellar 
luminosity during the MS phase and limits the
redwards motion in the HR diagram. 

The widening of the MS produced by rotation mimics the effect of an overshoot
beyond the convective core
(see Talon et al. \cite{Tal97}, paper VII). Since the observed width of the MS has 
often been taken to parameterize the size of the convective core, 
the above argument
shows that rotating models tend to decrease the amplitude of the overshoot
necessary to reproduce the observed MS width.

\begin{figure}[tb]
  \resizebox{\hsize}{!}{\includegraphics[angle=0]{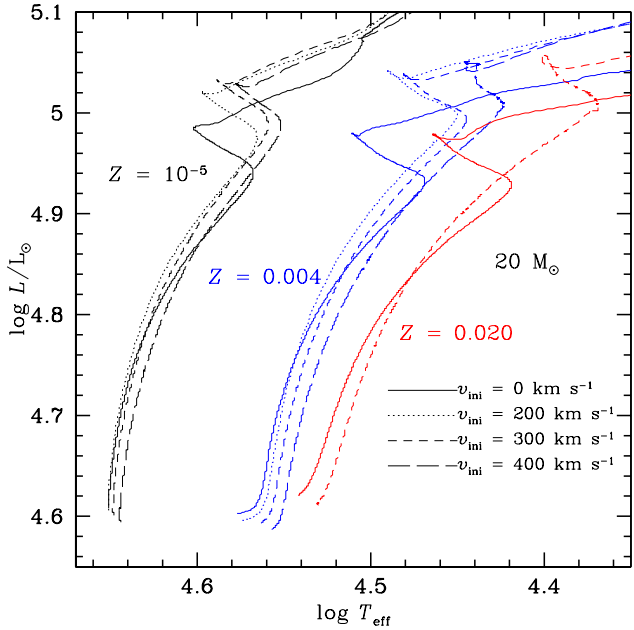}}
  \caption{Evolutionary tracks for rotating 20 M$_\odot$ models with different
initial velocities and various initial metallicities. 
The initial velocities $\upsilon_{\rm ini}$ are indicated. 
See Table 1 for more details on the models at $Z = 10^{-5}$.}
  \label{fv9v20}
\end{figure}

Figure~\ref{fv9v20} shows the 
evolutionary tracks of 20 M$_\odot$ models for different initial
velocities and metallicities during the H--burning phase.
One sees that the extension of the MS due to rotation decreases
when the metallicity decreases. This results from the smaller increase of
the He--core due to rotation at low $Z$.
Typically, at the end of the MS at $Z = 10^{-5}$, the helium core mass in the
rotating 20 M$_\odot$ model 
($v_{\rm ini}$ = 300 km s$^{-1}$) is greater
by 11\% with respect to its value in the non--rotating model. The corresponding
increase at Z=0.004
is 23\%. The reason for this difference is the following one.
 When rotation brings fresh hydrogen fuel, in
the core, it brings also carbon and oxygen which act as catalysts in the CNO
burning. These
catalyst elements are of course in much lower abundances at $Z=10^{-5}$ than
at
$Z$ = 0.004 and thus the core enhancement is less pronounced.

At $Z$ = 0.004, the rotating star models with initial masses between 
9 and $\sim$25 M$_\odot$, are evolving, after the MS
phase, much more rapidly
toward the red supergiant stage (RSG) than the non--rotating models, in
agreement
with the observed number ratio of blue to red supergiants in the Small
Magellanic Cloud cluster
NGC 330  (Maeder \& Meynet \cite{MMVII}). 
As was extensively discussed, the balance between the blue and the red 
is very sensitive to many effects.  
At the  very low metallicity considered here, rotation does not
succeed in producing red supergiants, at least for the range of initial rotational velocities explored. 
The dominant reason appears to
be the smaller He--cores produced at lower $Z$ and the fact that the
growth of this core due to rotation is smaller than at higher $Z$.
In terms of the discussion by Maeder \& Meynet (\cite{MMVII}),
this reduces the central potential enough to keep a blue location
during the whole He--burning phase, and the amount of helium diffused
in the region of the shell is unable to compensate for the smaller core.
 Only, during the 
very last stages in the C--burning phase, when the core heavily contracts,
 does the central  potential grow enough to produce a red supergiant.


The rotating models do not present any well developed blue loop, except in the
case of the 2 M$_\odot$ model. 
In that respect the situation is similar to the case of the non--rotating
models (see Sect. 5 above). Interestingly, we note that the rotating models,
with initial mass below 7 M$_\odot$, evolve redwards during the AGB--phase.
This a consequence of the third dredge--up, which brings at the surface carbon 
and oxygen synthesized in the He--burning shell, as well as primary nitrogen 
built up in the H--burning shell (see Sect. 7 and 8). The important enhancements of 
these elements at the surface make the star to behave as a more metal 
rich star and thus push it to a redder location in the HR diagram.

In the present
grid no model enters the Wolf--Rayet phase. At the end of the C--burning phase,
the mass fraction of hydrogen at the surface 
of the rotating 60 M$_\odot$ model is still important ($\sim 0.48$), although
much lower than at the surface of the
corresponding non--rotating model ($\sim$ 0.77). It is likely that
 more massive or faster rotating star models may enter the Wolf--Rayet 
phase before central He--exhaustion.

\subsection{Masses and mass--luminosity relations}

Table~\ref{tbl-1} presents some properties of the models. Columns 1 and 2 give
the initial mass and the initial velocity $v_{\rm ini}$ respectively.
The mean equatorial rotational velocity $\overline{v}$ during the MS phase is
indicated in column 3.
The
H--burning lifetimes $t_{\mathrm{H}}$, the equatorial velocities $v$, the
helium surface abundance $Y_{\mathrm{s}}$ and the 
surface ratios N/C and N/O at the end of the H--burning phase and normalized to
their initial values are given in columns 4 to 8.
The columns 9 to 13 present some properties of the models 
at the end of the core He--burning phase; $t_{\mathrm{He}}$ is the He--burning
lifetime. Some characteristics of the last computed models are given in columns
14 to 18; $M_{\rm fin}$ is the final stellar mass. 
For stars with initial mass superior or equal to 9 M$_\odot$, the
final stage corresponds to the end of the
C--burning phase.
For the lower initial mass stars, it corresponds
to the beginning of the Thermal Pulse AGB (TP--AGB) phase. Typically
the rotating 3 M$_\odot$ model was computed until the fifth 
thermal pulse. For the intermediate mass stars, the mass
fractions
of carbon (C), oxygen (O) and nitrogen (N) at the surface of the stars are
given.

Rotation, by enhancing the luminosity and lowering the effective gravity,
increases the mass loss rates (Maeder \& Meynet \cite{MMVI}).
As a consequence, the final masses of the rotating models are smaller. At the
metallicity $Z = 10^{-5}$,
except for the 60 M$_\odot$ model, the effects of rotation on the final stellar
masses are very weak
(see Table~\ref{tbl-1}). 


In general, rotation makes the star overluminous for their actual masses.
Typically 
for $v_{\rm ini}$ = 300 km s$^{-1}$, the luminosity  vs. mass (L/M) ratios at
the end
of the MS are increased  by 10--14\% for stars in the mass range from 3 to 40
M$_\odot$. 
This results essentially from the He diffusion in the radiative envelope which
lowers the
opacity and makes the star overluminous.
In the 60 M$_\odot$ model, mixing is particularly efficient and 
the increase of the L/M ratio amounts to 23\%. 
For the 2 M$_\odot$ model, the L/M ratio
is decreased in the rotating model, by 6--7\% . In this last case, the
convective core during the H--burning phase
disappears very early, when the mass fraction of hydrogen in the center is still
high ($X_{\rm c}$ = 0.52 in the
$v_{\rm ini}$ = 300 km s$^{-1}$ model). This puts farther away from the surface
the region where helium is produced and thus
slows down the helium diffusion in the outer envelope. 

The increase of L/M, due to rotation, at $Z =10^{-5}$ are in general inferior to those obtained
at $Z$ = 0.004, 
which are between 15--22\% (Maeder \& Meynet 2001). This is mainly a
consequence
of the following fact: at $Z=10^{-5}$, the increase of the H--burning convective
core due to rotation
is inferior to that obtained at $Z$ = 0.004 for the same value of $v_{\rm ini}$.

\subsection{Lifetimes}

Generally we can say that
 the MS lifetime duration is affected by rotation at least through
three effects:
\begin{itemize}

\item 1) Rotation increases the quantity of hydrogen burnt in the core. 
This increases the MS lifetime.

\item 2) The hydrostatic effects of rotation make a star of a given initial mass
to behave
as a non--rotating star of a smaller initial mass. This tends to increase the
MS lifetime.

\item 3) Rotation increases the helium abundance in the outer radiative
envelope. This tends
to make the star overluminous with respect to its non--rotating counterpart and
thus
to reduce the MS lifetime.

\end{itemize}
When the metallicity decreases, the effect number 3 tends to
become the most important one.

 
This can be seen 
from a detailed comparison of the tracks in Fig.~\ref{fv9v20}. Indeed
the evolutionary tracks for our rotating 20 M$_\odot$ models 
($v_{\rm ini}$ = 300 km s$^{-1}$) become overluminous with
respect to the non--rotating tracks at an earlier stage for
lower metallicities.
This mainly results from the fact that
when the metallicity decreases, rotational mixing is more efficient
(see Sect. 3).
Also, at lower $Z$, the stars are more compact and
therefore the timescale for mixing, which is proportional 
to the square of the radius, decreases. This  favours
the helium diffusion in the outer envelope.
The diffusion of hydrogen into the core, which would increase the MS lifetime,
is not really affected, because
hydrogen just needs to migrate over the convective core 
boundary to be engulfed into the core.
One notes also that for given values of the equatorial velocity
and of the initial mass,
the ratio  $\Omega/\Omega_{\rm c}$ of the angular velocity to the break--up velocity  
decreases whith the metallicity. Thus
the hydrostatic effects,
which usually make the star fainter,
are  in general smaller at lower $Z$.

As a consequence of the above effects,
at $Z=10^{-5}$, the MS lifetimes are 
decreased by about 4--14\% for the mass range between 
3 and 60 M$_\odot$ when   $v_{\rm ini}$
increases from 0 to 300 km s$^{-1}$ (cf. Table~\ref{tbl-1}). 


We notice that the 
 rotating 2 M$_\odot$ model has  a longer MS phase than its non--rotating counterparts,
in contrast with what happens for higher initial mass stars.
This is because,
when the initial mass decreases, the hydrostatic effects
become more and more important (see Fig.~\ref{HR}).

For what concerns the effects of rotation on the He--burning lifetime, let us simply say that  when   $v_{\rm ini}$
 increases from 0 to 300 km s$^{-1}$,
the changes in the He--burning 
lifetimes are inferior to 10\%. 

\section{Evolution of the chemical abundances at the surface}

\begin{figure}[tb]
  \resizebox{\hsize}{!}{\includegraphics[angle=0]{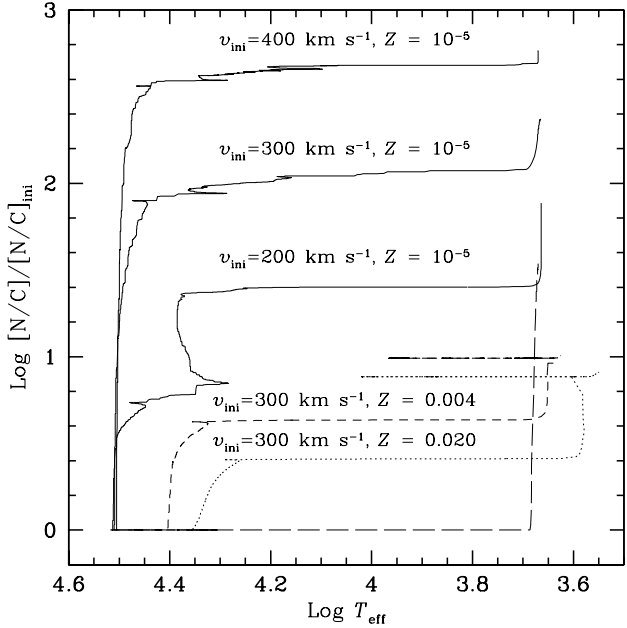}}
  \caption{Evolution as a function of $\log T_{\rm eff}$ of the abundance
ratio N/C where N and C
are the surface abundances of nitrogen and carbon
respectively. The abundance ratios are normalized to their initial values. The
tracks
are for 9 M$_\odot$ for different values of the metallicity, $Z$, and rotation.
The long--dashed line, at the bottom, corresponds to a non--rotating
9 M$_\odot$ stellar model at $Z=10^{-5}$.
}
  \label{M9NC}
\end{figure}

  Fig.~\ref{M9NC}
shows the evolution of the N/C ratios  in models of
rotating stars with 9 M$_{\odot}$ for the
initial $Z$ = 0.02, 0.004 and  10$^{-5}$. Values for other stellar
masses may be found in  Table~\ref{tbl-1}.

At zero rotation, for any $Z$ and any masses there is no enrichment during 
the MS phase (except at $Z$=0.02 for M $\geq 60 M_{\odot}$ due to very high
mass loss). At 9 M$_{\odot}$  for an initial
rotation of 300 km s$^{-1}$, 
we notice an increase of the N/C ratio already during the
MS phase. In fact most of the increase in N/C 
is in general  built during
the MS phase, and this results from the steeper $\Omega$--gradients
and greater compactness. 
The \emph {relative} growths of the N/C ratio
do not change very much from models with $Z$ = 0.02 to models 
with $Z$ = 0.004, however there is an increase by two orders of a magnitude
for   $Z$ = 10$^{-5}$. Of course this large N/C enhancement is accompanied
by a small enrichment in helium at the surface, 
typically of a few hundredths as shown by Table~\ref{tbl-1}. 
Fig.~\ref{M9NC} illustrates the fact we find throughout this
work, i.e. that the various  effects  of rotation on the 
internal structure, the surface 
composition and the yields are in general much higher at lower metallicities.

In  Fig.~\ref{M9NC}, we notice for  $Z = 10^{-5}$ an   increase by a factor 
4.5 of the N/C ratio for an increase of 100 km s$^{-1}$ of the initial
rotation. As illustrated by Table~\ref{tbl-1},
 during the He--burning phase the rotation
velocities become all the same whatever the initial
rotational velocities. Thus, in the He--burning
phase we  may have very different surface chemical compositions
for actually similar rotation velocities. This is likely true for 
all stellar masses where fast rotation is present, but the effect
is in general larger for larger masses.

Curiously enough, at very low $Z$ the fast rotating stars
of intermediate masses  which reach the TP--AGB phase
(this occurs for M $\leq 7 {\rm M}_{\odot}$) 
get a higher $Z$ during this phase due to their enrichment in 
CNO elements. As an example, a  7 M$_{\odot}$ has 
a  X(CNO) = 3.1 $\cdot 10^{-3}$ which is 430 times
the initial CNO content. Thus very low $Z$ stars
may become higher $Z$ stars near the end of their 
evolution. This might also affect the composition of planetary nebulae
in low $Z$ regions.

\section{Primary nitrogen production}

\subsection{ Brief recalls on the nitrogen synthesis}

Nitrogen 
is mainly produced in the CN branch of the CNO cycles
within H--burning stellar zones (see Clayton \cite{cl83}; Arnett \cite{arn}).
More precisely, three reactions occur to transform $^{12}$C to
$^{14}$N: $^{12}$C$(p,\gamma)^{13}$N$(\beta^+,\nu)^{13}$C$(p,\gamma)^{14}$N.
Nitrogen can also be produced in the ON cycle by transformation 
of $^{16}$O, but at a much slower rate. The reaction $^{14}$N$(p,\gamma)^{15}$O
which depletes nitrogen has a relatively low cross 
section enabling $^{14}$N to accumulate with time.
Thus, $^{14}$N  is usually  the daughter element, hence a secondary element, of
the CNO  initially present in stars. 

Nitrogen is said to be primary (Talbot \& Arnett \cite{tal}),
 if it is formed in a 
star not from the initial CNO elements, but from the
hydrogen and helium.  Of course, the reactions forming 
primary nitrogen are those mentioned above, but the 
sequence of events is different: the formation of primary
nitrogen implies firstly the synthesis of some $^{12}$C
by the 3$\alpha$--reaction in a helium burning region, then this 
new $^{12}$C  needs to be transported in an hydrogen
burning region, where the CNO cycle will convert it
to nitrogen. Thus, primary nitrogen is likely to be formed
in stars with a He--burning core  and a CNO burning shell,
provided there is some transport mechanism between the two.
The absence of such transport is the main reason why
current models do not produce in general any primary nitrogen.

If  $^{14}$N is of primary origin,  the $^{14}$N--abundance is proportional to
that of the  other primary heavy elements.
While if nitrogen is secondary, the increase in the abundance 
of $^{14}$N  is proportional to the initial CNO content and 
thus in the chemical history of a galaxy the $^{14}$N--content 
will be proportional to the square of the CNO and metal content. 
These different behaviors provide the basic test for
ascertain the primary or secondary origin of nitrogen.

 The observations point toward
the need of primary nitrogen sources at low metallicities
(see below). The main problem is that the  stellar models,
 unless ad hoc hypothesis
are made, do not  currently predict any primary nitrogen.
 This suggests that some physical process 
may be missing in the stellar models.

There are two other related problems. 
At solar metallicities, the observations do not suggest the
production of primary nitrogen. Thus, a global question
is how is changing the respective efficiencies of the primary
and secondary $^{14}$N productions during the evolution of galaxies. 
The other question concerns 
 the relative importance of massive and intermediate mass stars
in the production of primary and secondary nitrogen.
This point is important in relation for the
interpretation  in terms of  the star formation history
the N/O ratios observed at high redshifts
 (Pettini et al. \cite{Pett95}; Lu et al. \cite{Lu98}).
Nitrogen, primary and secondary, is produced in the longest 
and main evolutionary phases. As long as its production is
not well understood, we may doubt of the correctness 
of  the models
for these main phases of stellar evolution.

\subsection {The observations of the N/O ratio}.

There are several kinds of evidences in favour of primary nitrogen
in the early phases of  the evolution of galaxies.

--1. An indication of primary nitrogen is provided by
the study of the N/O ratio in low metallicity stars of
 the galactic  halo.  The ``discovery'' of primary nitrogen 
was made by Edmunds \& Pagel (\cite{edm}) in a study of the N/O ratio
 in such stars and in some external galaxies. Following this 
work, others authors  (Barbuy \cite{bar};
 Tomkin \& Lambert \cite{tom}; Matteucci \cite{mat};
 Carbon et al. \cite{car}; Henry et al. \cite{Ha00}) have shown that the
 ratio N/O of  nitrogen to oxygen remains constant with a plateau at log N/O
$\simeq$ -1.7  in the early evolution of the Galaxy, thus implying a primary
origin of nitrogen. The limit in metallicity above which secondary 
production of nitrogen becomes important is difficult to
fix with precision, since the transition is progressive. It
is around  $12+ \log$ O/H = 7.8 to 8.2 according to Henry et al.
(\cite{Ha00}; cf. also Izotov \& Thuan \cite{izo99}). Since for the Sun
 one has $12+ \log¨$ O/H =8.9, this means at a metallicity $Z$ equal
to $Z_{\odot}/12$ to  $Z_{\odot}/5$. Above this limit, 
the N/C and N/O ratios grow rapidly, implying
that nitrogen is essentially a secondary element. It is not 
known whether the primary production stops completely
for $Z$ values higher than the above limit.
A good way to check it would be to measure 
the sum of CNO elements in planetary nebulae 
of the SMC, LMC
and Galaxy, to see whether this sum is higher than the initial
local CNO content of these galaxies.
 
--2. A very compelling evidence for primary 
$^{14}$N is provided by the study
 of the  N/O ratios in ionized HII regions of blue compact
 dwarf galaxies (Thuan \& Izotov \cite{thu}; 
Kobulnicky \& Skillman \cite{kob96};
Izotov \& Thuan \cite{izo99}; Izotov \& Thuan \cite{izo00}).
These HII regions also show a plateau of N/O at log N/O 
$\simeq$ -1.7
below $12+ \log {\rm O/H} \simeq 8.0$, while above this limit
the N/O  ratio is a steeply growing function of O/H,
as due to the secondary production of nitrogen.
  An example of a low metallicity galaxy is  IZw~18, which has the 
lowest known metallicity (1/50 of solar), and which shows 
indications of primary nitrogen (Kunth et al. \cite{kun95}; 
Izotov \& Thuan \cite{izo99}).
The study of the N/O ratio in spiral galaxies by van Zee et al.
(\cite{vanZee98})  well  confirms the same result,
with the difference that the authors find a plateau 
below  $12+ \log$ O/H = 8.45, i.e. for abundance of heavy elements
less than 1/3 solar.

A problem was that some low $Z$ damped Ly$\alpha$ systems
 have N/O ratios lower than those observed in the  HII regions 
 of blue compact dwarf galaxies of the same $Z$ (Pettini et al. 
\cite{Pett95}). However, the apparent discrepancy  has been
 resolved  by models of damped Ly$\alpha$  systems which account for both
ionized  and neutral regions (Izotov et al. \cite{izo01}).

--3. Another argument for primary nitrogen production comes from the
observed gradient of  N/O in spiral galaxies.  If nitrogen 
is purely a secondary element, the
N/O gradient should be identical to that of O/H. 
In general, the N/O gradients tend to be shallower than the 
O/H gradients (Vilchez \& Esteban \cite{Vil96}). The various data
on the galactic gradients of N/O (Rudolph et al. \cite{Rud97};
Garnett et al. \cite{Gar97}; Ferguson et al. \cite{Fer98}; 
Henry \& Worthey \cite{Hen99}) 
generally show 
 that the N/O gradient is  relatively flat at low metallicity $Z$,
which supports the conclusion that  the production of nitrogen is dominated by
primary processes at low $Z$ , while at solar or higher $Z$
the similarity of the N/O and O/H gradients support the view that
nitrogen is secondary.

The situation is rather confused concerning the masses of 
the stars responsible for the injection of primary nitrogen.
There are authors supporting the origin of primary 
 nitrogen in massive stars (Matteucci \cite{mat}; Thuan \& Izotov 
\cite{thu}; 
Izotov \& Thuan \cite{izo99}; Izotov \& Thuan \cite{izo00}).
Their main argument is the low scatter of the observed N/O ratios at low $Z$.
Indeed, if nitrogen is synthesized in massive stars, there is no time delay
between the injection of nitrogen and oxygen and thus a rather small scatter
would result. On the contrary, if the primary nitrogen is made in intermediate
mass stars, the N/O ratio increases with time, since these stars release their 
nitrogen much later than 
massive stars do eject their oxygen.
 This would lead to a larger scatter in the observations,
because  galaxies are observed  at various stages of their evolution. 
Izotov \& Thuan (\cite{izo99}) suggest also that because of the 
intermediate mass star delay, the faster evolving massive stars
must be a significant source of primary nitrogen in order
 to raise the $\log$ (N/O) ratio to the observed plateau
 at 12+$\lg({\rm O/H}) \sim 7.2$, a metallicity they assume
to correspond to a 
galactic age too short to allow nitrogen ejection by the intermediate 
masses.

 The situation may be not so clear, because
 some  studies found that a  significant scatter does exist
 (Garnett \cite{Gar90}; Skillman et al. \cite{Ski97}). 
Also, Henry et al. (\cite{Ha00}) have calculated chemical evolution models 
which  support the view that intermediate mass stars 
between 4 and 8 M$_\odot$, with an age of about 250 Myr, 
are likely to dominate the nitrogen production.

 What can we deduce if we accept the fact that the N/O
 values show a great scatter at fixed value of O/H ?
 A possibility might be that the observed scatter
occurs because we are  observing a large sample of HII regions 
in various stages of oxygen and
nitrogen enrichments. This picture implies that most observed data
should have relatively high N/O values with fewer points, representing
those objects experiencing sudden oxygen enrichment, located below the 
bulk of data,  since presumably bursts are followed by 
relatively long periods of quiescence,
with relatively higher N/O ratios.

The reality looks different.
 The distribution of points in N/O vs. O/H plane
 reveals that most points seem to be clustered at
relatively low values.  This suggests that the ``equilibrium'' 
or unperturbed locus where most HII
regions reside is the low N/O envelope. 
Thus, this suggests that the excursions caused by 
sudden injections of material are actually upward,
 toward the region of fewer points. This picture seems
consistent with the lack of evidence for localized oxygen
 contamination from massive stars in H II regions 
(Kobulnicky \& Skillman \cite{kob97}).
The falloff in points above the N/O envelope 
is more consistent with injections
of nitrogen rather than oxygen. 
In this case, the nitrogen source might be WR stars
or luminous blue variable stars, both of which were
 considered by Kobulnicky et al. (\cite{kobSR97})
 in their study of nitrogen--enriched H II regions in NGC 5253.
 They expect also a simultaneous enrichment 
in helium, and thus H II regions exhibiting high 
values of N/O should also be checked
for evidence of helium enrichment.

\begin{figure*}[tb]
  \resizebox{\hsize}{!}{\includegraphics[angle=0]{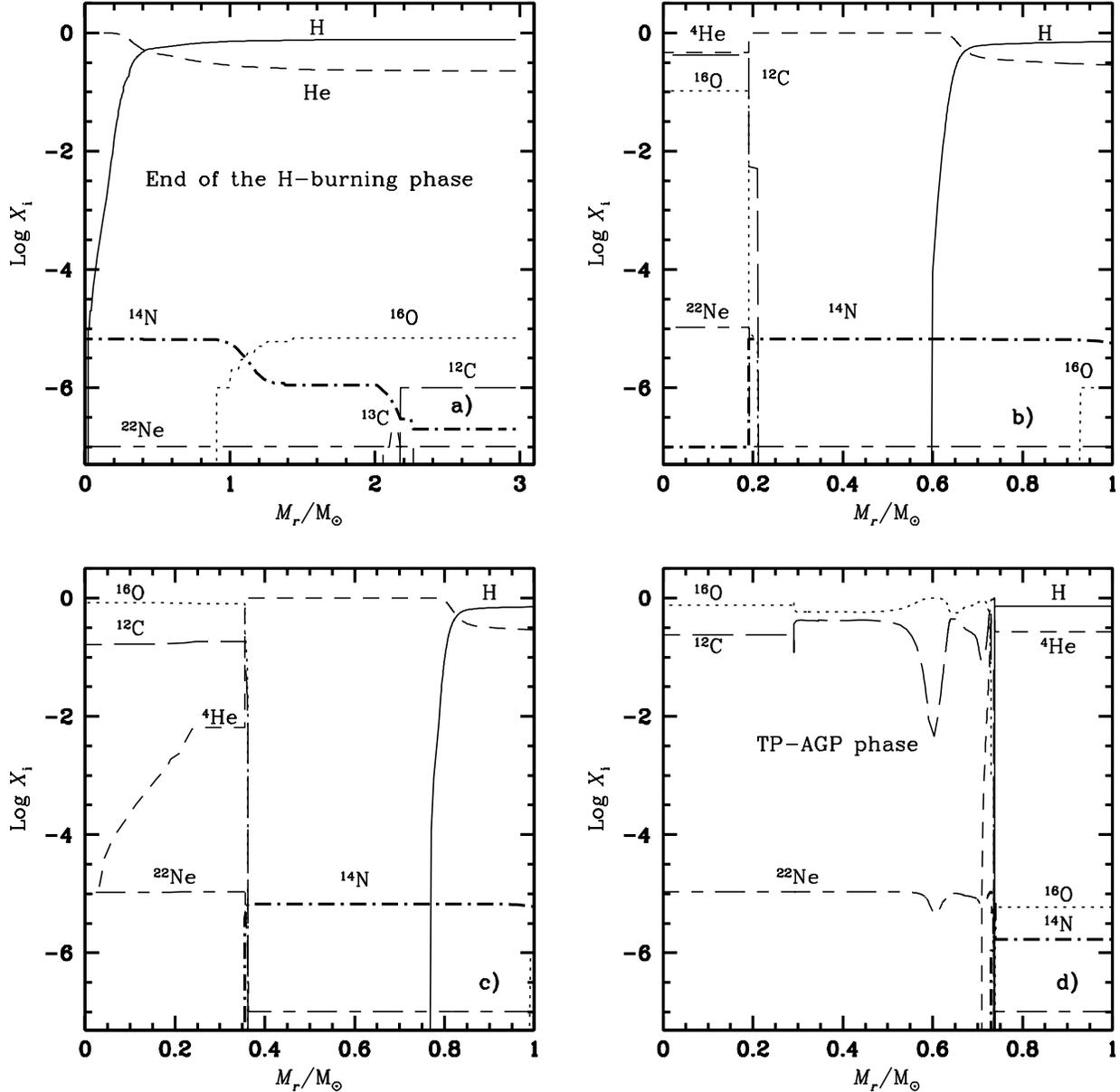}}
  \caption{Variations as a function of the langrangian mass coordinate $M_{r}$ 
of the abundances of various elements inside a non--rotating 3 M$_\odot$ model at the metallicity $Z = 10^{-5}$. 
Panel a) shows the chemical structure at the end 
of the core H--burning phase. Panels b) and c) at the middle and at 
the end of the core He--burning phase. 
The structure after one pulse along the Thermal Pulse--AGB phase is shown on panel d).}
  \label{reu0}
\end{figure*}

\begin{figure*}[tb]
  \resizebox{\hsize}{!}{\includegraphics[angle=0]{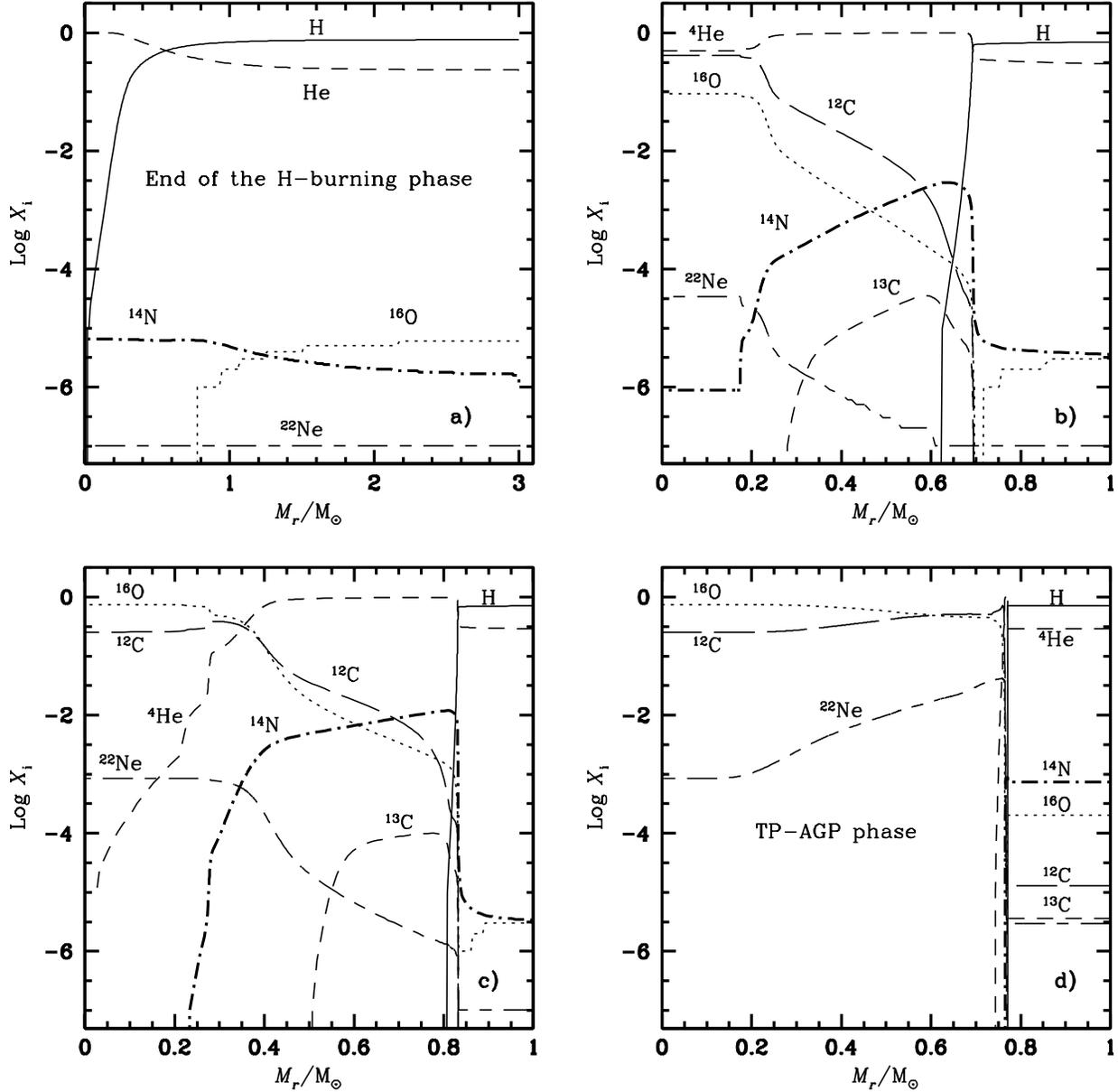}}
  \caption{Same as Fig.~\ref{reu0} for a rotating 3 M$_\odot$ model at the metallicity $Z = 10^{-5}$. 
The initial velocity on the ZAMS is 300 km s$^{-1}$, which corresponds to an
average surface equatorial velocity during the Main Sequence equal to 
$\sim$230 km s$^{-1}$. Panel a) shows the chemical structure at the end 
of the core H--burning phase. Panels b) and c) at the middle and at 
the end of the core He--burning phase. 
The structure after the first five pulses along the Thermal Pulse--AGB phase is shown on panel d).}
  \label{reu1}
\end{figure*}

\begin{figure*}[tb]
  \resizebox{\hsize}{!}{\includegraphics[angle=0]{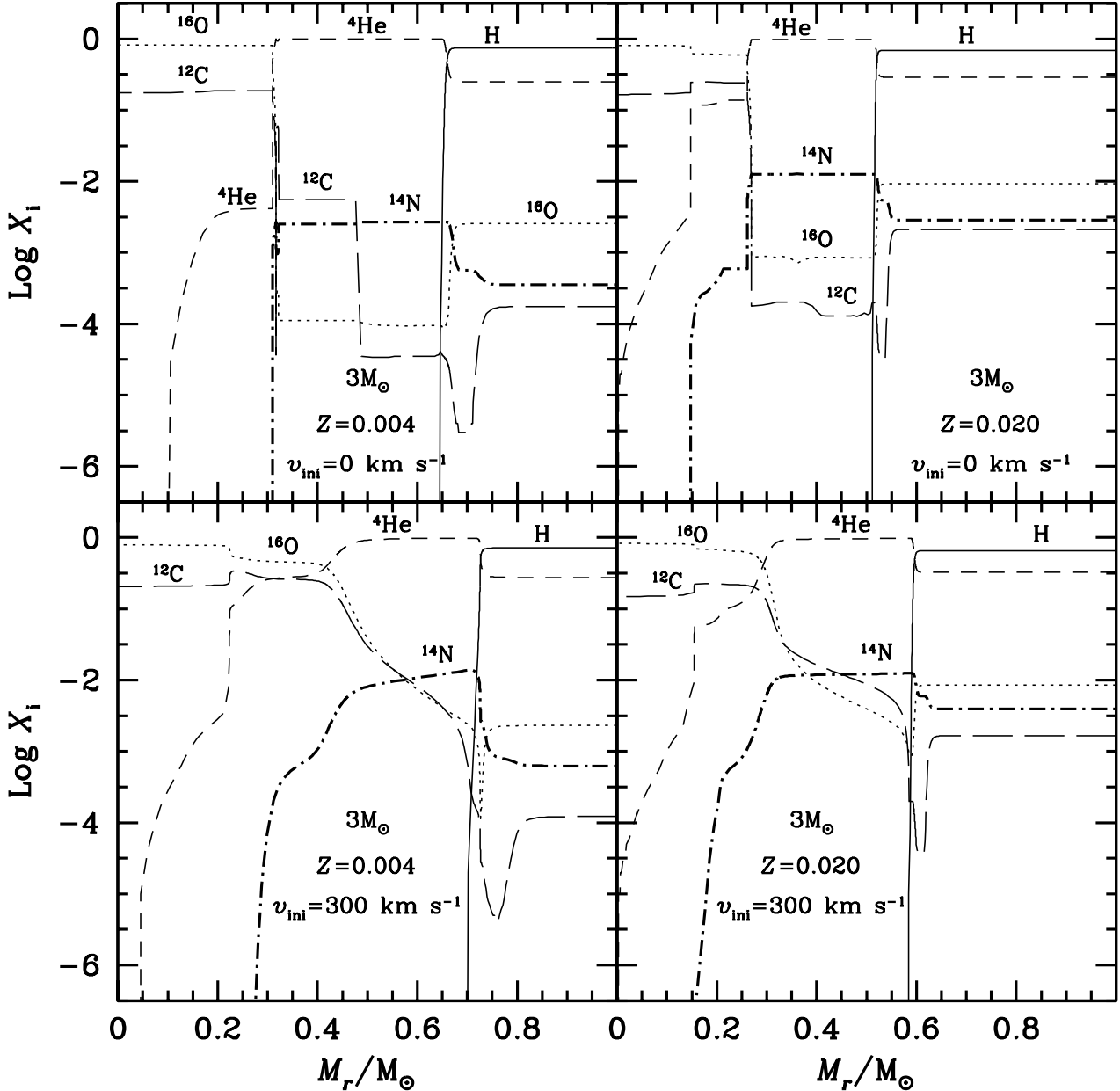}}
  \caption{Variation of the abundances of some elements in the intershell region
of 3 M$_\odot$ models 
at the end of the He--burning phase for the metallicities $Z$ = 0.004 and 0.020.
The initial velocities $v_{\rm ini}$ are indicated.}
\label{N1432Z}
\end{figure*}

\begin{figure*}[tb]
  \resizebox{\hsize}{!}{\includegraphics[angle=-90]{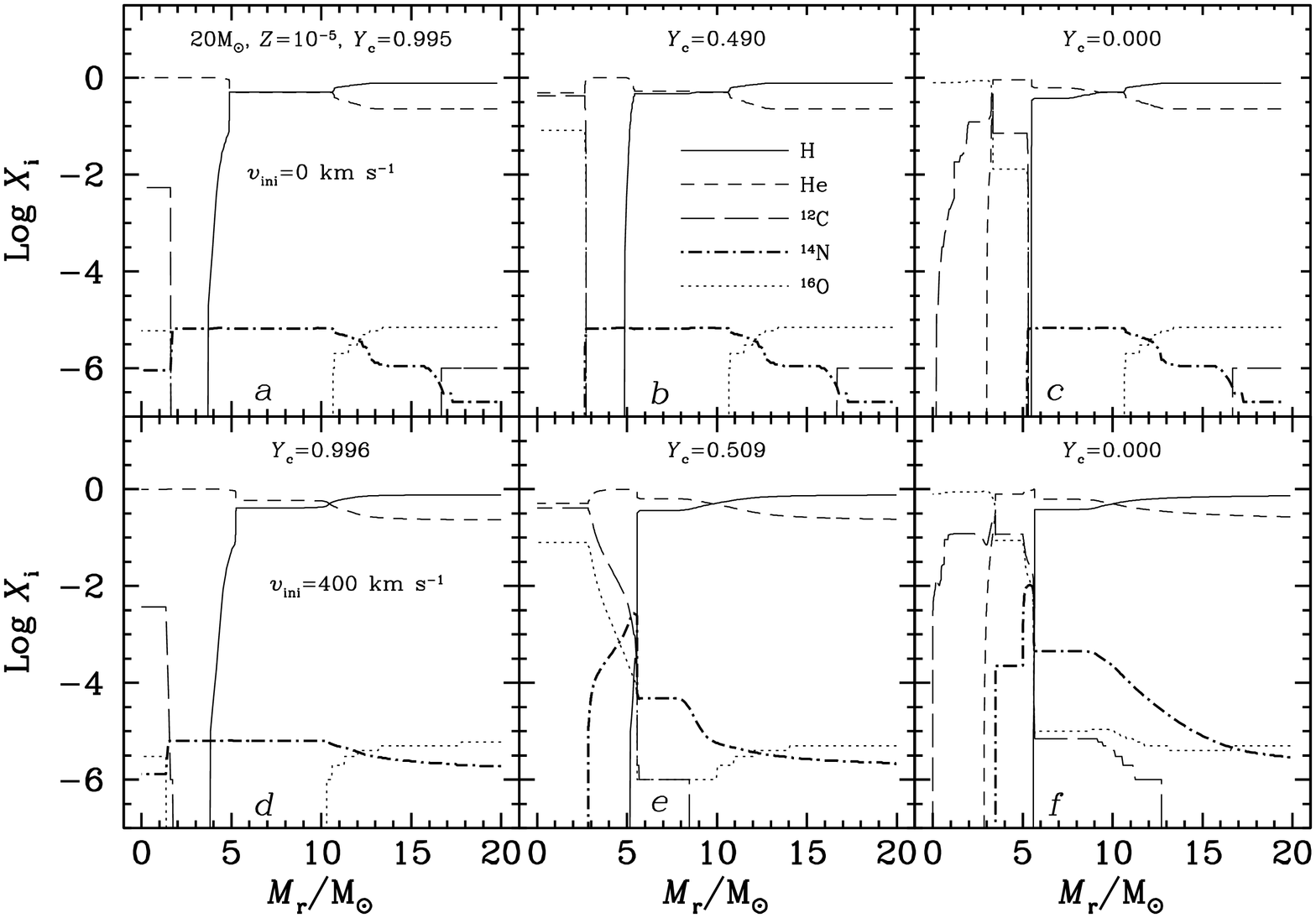}}
  \caption{Variation of the abundances of some elements inside 20 M$_\odot$
models at various stages during the
core and shell He--burning phases. The initial metallicity is $Z=10^{-5}$. The
upper panels refer to non--rotating models.
Panel {\it a} corresponds to the beginning of the core He--burning phase, panel
{\it b} shows the situation
at the middle of the He--burning phase ($Y_c \sim$0.50) and panel {\it c} at the
end of the C--burning phase. Panels {\it d} to {\it f}
show the same stages for the corresponding rotating models with $v_{\rm ini}$ =
400 km s$^{-1}$.}
  \label{N1420}
\end{figure*}


\subsection{The existing stellar models}

For massive stars,  there is at present  no model
producing primary nitrogen unless some ad hoc assumptions are made
 in order to reproduce the
observed N/O ratio at low metallicity (Timmes et al. \cite{tim}).
 In these ad hoc models, some
mixing is permitted between the helium-- and hydrogen--burning zones.
Some  primary nitrogen may also be produced in low--metallicity
 massive stars via some adjusted convective overshoot
(Woosley \& Weaver \cite{WW95}).
Without any physical explanation, it is difficult to understand why the
production of primary nitrogen only occurs at low metallicities.
 Models of metal free Population III stars (Umeda et al. \cite{ume})
 produce some primary nitrogen, but in too low quantities 
to reproduce the observed plateau (see also Heger et al. \cite{heal200}).

There is an extensive literature on AGB star models (see for example
Forestini \& Charbonnel \cite{FC97}; Boothroyd \& Sackmann 
\cite{BS99}; Marigo \cite{mar98}, \cite{mar01}).
Up to the phase of thermal pulses on the  AGB  branch, 
the intermediate mass star models predict no primary nitrogen
production. Only when the star enters the phase of thermal pulses,
some He--burning products may be transported into the 
H--burning shell, thus producing  some primary nitrogen.
These models are complex and require a lot of computing time.
This is why the AGB models
(Renzini \& Voli \cite{RV81}; Marigo \cite{mar98}) 
are ``synthetic'' models, which means that the model 
parameters follow some analytical relations that have 
generally been fitted to the observations (this is the case
for example for the minimum stellar mass experiencing 
the third dredge--up). In addition the dependence of 
this minimum mass on metallicity is based on observation. The same
kind of adjustments are made for the occurrence of the hot bottom burning.
 While this may
 be useful for some purposes, it cannot be claimed 
that it represents consistent physics  leading to primary 
nitrogen production. 
The  published stellar yields  
for intermediate mass stars are generally  based on such
 synthetic models.
 According to the models of Marigo (\cite{mar01}),
 the primary nitrogen production depends heavily
on the parameters describing the hot bottom burning 
and the third dredge--up, both processes which are not adequately
described in complete stellar models. 
This means that for intermediate mass stars the
 primary nitrogen production
is not a fully consistent output. 

For completness, we also mention here that some explanations 
of the N/O ratios advocates galactic processes, such as
 differential outflows of the chemical elements 
produced by galactic winds (Edmunds \cite{edm90}).
Oxygen is predominantly made in high--mass stars that
undergo more violent explosions than intermediate mass stars,
 thus is more likely to be removed
from the galaxy. This differential outflow results
in a decrease in the effective yield
for oxygen with a corresponding increase in the N/O ratio.
By observing massive spiral galaxies, 
van Zee et al. (\cite{vanZee98}) tried to minimize the 
complicating effects of gas outflow/inflow. 
They performed nitrogen and oxygen 
abundance measurements for 185 H II regions spanning a range
of radii in 13 spiral galaxies and obtained for the N/O ratios
the same behavior as in low--metallicity 
dwarf galaxies. This result suggests that the observed trend 
in dwarf galaxies is not
due to the outflow of enriched material in a shallow gravitational potential.
They conclude that low--metallicity H II regions in all types of galaxies do
show evidence of primary nitrogen production.

\subsection{The physics of the production of primary nitrogen
in low $Z$ rotating models}

We need to look with some details the physics which
 determines
the synthesis of primary nitrogen and more generally the 
particular yields at low $Z$. Some effects have
already been examined by Meynet \& Maeder (\cite{MMlettreN}).
 We organize the discussion in a
systematic way:\\
-- 1. Effects of rotation in a 3 M$_{\odot}$ model at $Z$ = 10$^{-5}$.\\
-- 2. Same problem at $Z$ = 0.004 and 0.02.\\
-- 3. Effects of rotation in a 20 M$_{\odot}$ at $Z$ = 10$^{-5}$, 0.004\\
\hspace*{7mm} and 0.020.

-- 1. Figs.~\ref{reu0}  and ~\ref{reu1} compare the 
variations of the abundances inside a non--rotating 
and a rotating 3 M$_\odot$ model with $Z =10^{-5}$ at various 
evolutionary stages.
At the end of the H--burning phase (panels a), we notice 
the milder $\mu$--gradient at the very edge of the core, 
this contributes to make 
slightly larger 
He--cores in rotating models. This  characteristic 
is generally larger in larger masses,
it also  persists and increases
 in  further stages. We notice a significant
 diffusion of He and N throughout the star in the
rotating models. At the middle and at the end of 
the He--burning phase (panels b and c),
the differences in the chemical profiles are striking. 
In the non--rotating case, there is no new $^{12}$C 
outside the convective core,
and therefore there is no primary $^{14}$N produced.
While in the rotating model,
 $^{12}$C (together with some $^{16}$O ) is 
diffusing out the He--burning core and 
when it reaches the H--burning shell,
it is turned  by the CNO--cycle into primary $^{14}$N, 
producing a big bump of $^{14}$N
and a smaller one in $^{13}$C.
 As the H--shell migrates toward the exterior, the
bumps of  primary $^{14}$N  and  
$^{13}$C also progressively extends toward
 the exterior.  The height of these two
bumps is growing during the He--burning phase,
 since diffusion is bringing more and more  $^{12}$C 
which is turned to $^{14}$N, this explains the growth
of the peak in $^{14}$N at the outer edge of the intershell
region. The abundance of $^{14}$N in the rest of the
intershell region is also growing with time and this
is likely due to  the inward diffusion of nitrogen from the 
peak. Typically, at the end
of the He--burning phase, the $^{14}$N--abundance in the
intershell zone has
increased by  3 orders of a magnitude  with respect 
to $^{14}$N  in the corresponding non--rotating
model. At this stage, 
the integrated quantity of new nitrogen synthesized 
is 3.22 $\cdot$ 10$^{-3}$ M$_\odot$,
while it is only 3.32 $\cdot$ 10$^{-5}$ M$_\odot$ in the non--rotating model. 

In further stages, the He--burning shell progresses outwards, letting 
a  degenerate CO core behind it and transforming the
$^{14}$N into $^{22}$Ne (panels d in Figs.~\ref{reu0} and \ref{reu1}).
Also in the TP--AGB phase, the content in nitrogen of the outer convective
zone  increases a lot, due to both the facts that the outer convection
zone deepens in mass and that the diffusion at the base of the convective
zone  continues to proceed.
This is saving from destruction a large fraction of the primary 
nitrogen produced earlier. At the stage shown in Fig.~\ref{reu1}
nearly the whole quantity of $^{14}$N is in the outer convective envelope.
The integrated quantity of 
primary  $^{14}$N at this stage
is 1.58 10$^{-3}$ M$_\odot$, i.e. about 50\% of what was 
present in panel c) at the end of the He--burning phase. 
This fraction of 50\% does not change very much during the end of the
TP--AGB phase, because the edge of the CO--core and the outer envelope 
stay very 
close in lagrangian coordinates.
This nitrogen will be ejected 
by the AGB star either by the superwinds or in the planetary nebula. 

In the corresponding non--rotating model (see Fig.~\ref{reu0}, panel d),
we notice a very similar final structure with a large CO core surrounded
by a convective envelope and two thin shells at the basis of it.
As only differences, we notice the much smaller $^{14}$N and $^{16}$O abundances
in the envelope. Also, because
 central degeneracy is higher and thus there is more 
 cooling by neutrinos, the nuclear reaction  $^{12}$C($\alpha,\gamma)^{16}$O
 proceeds farther in the
outer core regions than in the inner regions, leading to a kick in  $^{12}$C and
a bump in $^{16}$O as observed in panel d) of Fig.~\ref{reu0}.

The production of $^{13}$C in the outer half of the 
 zone between the He--core and  the H--burning shell during the 
He--burning phase is relevant for the nucleosynthesis of 
``s--process'' elements, since $^{13}$C is an efficient neutron source.
The ``s--elements'' are produced when $^{13}$C is reached
by the outer progression of the He--burning shell. Clearly,
the formation of ``s--elements'' is strongly favoured in
rotating stars (see also Langer et al. \cite{lan99}).
Amazingly, in the convective envelope, the mass fraction of the CNO elements is
about 100 times the initial mass fraction of the heavy elements ! 

--2. It is interesting to compare the above results with those of models
of a 3 M$_{\odot}$ at higher metallicities. Fig.~\ref{N1432Z} shows 
models with and without rotation for $Z = 0.004$ and 0.020 at the
end of the helium burning phase, i.e. corresponding to  panel c)
in Figs.~\ref{reu0}  and ~\ref{reu1}.  As usual, the models with zero
rotation show flat curves separated by steep transitions due
to intermediate convective zones. The models with 
rotation at $Z$ = 0.004 and 0.020
show both the typical internal diffusion profile for $^{12}$C and $^{16}$O
outside  the core and noticeably the distributions of $^{14}$N
are rather similar to that observed for the models at $Z$ = 10$^{-5}$.
There is still some quantity of primary nitrogen in the intershell
model of the $Z$ = 0.004 model, but it is relatively negligible 
in the $Z$ = 0.02 model.
The maximum values of $^{14}$N in the interior are similar in the
three 3 M$_{\odot}$ models considered, independently of $Z$.

What are the reasons of this relative constancy ? For the models
illustrated, the central $T$ are about the same, as normal for 
He, C, O cores of about the same mass at the middle of the
 He--burning phase in rotating models. However, the temperatures at
 the basis of the H--burning shells
are different: $\log T = 7.573$ at  $Z$ = 10$^{-5}$ and 7.446 at $Z$ = 0.02.
This is consistent with the fact that the $Z$ = 10$^{-5}$  model is much
brighter (log $L$/L$_{\odot}$ = 2.787 compared to 2.013 at $Z$ = 0.02),
because of its much lower opacity. The
nuclear energy  production (mainly of the H--shell) necessary to supply
the stellar luminosity is adjusted, as usual,
by the temperature of the shell and not by the content in $^{14}$N.
The similarity of the distributions
 of $^{14}$N  at low metallicity essentially results from the rotational
transport of material from the core. We have seen in Sect. 3
that the $\Omega$--distribution during the MS evolutionary phase
is different for different $Z$. However, in later phases the $\mu$--contrast
between the dense core and  the surrounding layers is about the same
and this contrast determines the $\Omega$--gradient and  in turn the
importance of  the diffusion. Thus, the diffusion of $^{12}$C 
outside the core is not very different in models of
same mass and rotation, and as a consequence the 
same is true for  $^{14}$N. 

At $Z$ = 10$^{-5}$ the gradient of $^{14}$N between the H--shell
and the envelope is much larger than at $Z$ = 0.02, because the 
difference between the peak of $^{14}$N in the intershell region
and the cosmic abundance in the envelope is also much larger.
This has two consequences: firstly, the inward progression of
 the outer convective zone will bring relatively much more
$^{14}$N in the envelope; secondly
there is also more diffusion of $^{14}$N in the envelope
at very low $Z$.

Finally, we also emphasize (cf. Meynet \& Maeder \cite{MMlettreN})
that in the TP--AGB phase the distance between the He-- and the H--burning
shells  is much smaller in lower $Z$ models. This effect will certainly
influence considerably the occurrence and properties of the relaxation oscillations.
Also, this effect makes the transport of  $^{12}$C from the He--burning
shell to the H--burning shell much shorter, since the timescale
for diffusion varies with the square of the distance. In this respect,
the smaller intershell region in lower $Z$ models also favours
the  increase of the abundance of primary nitrogen
in the envelope. 

In summary, the higher production of primary $^{14}$N 
in very low $Z$ models results mainly from
the relatively stronger peak of primary $^{14}$N 
 built by rotational mixing
 in the intershell region during the He--burning phase,
a part of which is entering the envelope during its inward migration
and another part is brought to the envelope by the diffusion,
which is favoured by the smaller
 intershell region during the TP--AGB phase.

3) Let us examine the 20 M$_{\odot}$ models at $Z$ = 10$^{-5}$,
0.004 and 0.020. In Fig.~\ref{N1420} for  $Z$ = 10$^{-5}$, we see
for the rotating model
at the middle of the He--burning phase (panel e)
 the same kind of diffusion profile
of  $^{12}$C and $^{16}$O  outside the core, 
as in the corresponding 3 M$_{\odot}$ model. A similar,
although slightly smaller peak of primary $^{14}$N
is built between
the core and the H--shell. Contrarily to
smaller masses, where there is no central C--burning,
the intershell region remains large. The mainly primary $^{14}$N 
in this region will of course contribute to the yield,
as well as that in the outer envelope.
The abundance 
of $^{14}$N is increasing in the envelope during
He--burning and later phases. Since here,
contrarily to the low mass models, 
there is no inward migration of the
envelope, the increase of $^{14}$N in the envelope is only
due to the diffusion from the  $^{14}$N gradient in the 
H--burning shell. As in smaller masses, most of the 
$^{14}$N in the envelope is primary.

When we do a similar study in the 20 M$_{\odot}$ models
at $Z$ = 0.004, we notice that there is only a negligible 
amount of primary $^{14}$N produced and the abundance of $^{14}$N
in the envelope of the final models is  within a few percents
the same in the rotating and non--rotating cases.

In conclusion, we see that there is still some primary nitrogen
produced at $Z$ = 0.004 in intermediate mass models, as shown by
the 3 M$_\odot$ case, but nothing in the high mass stars.

\begin{figure}[tb]
  \resizebox{\hsize}{!}{\includegraphics{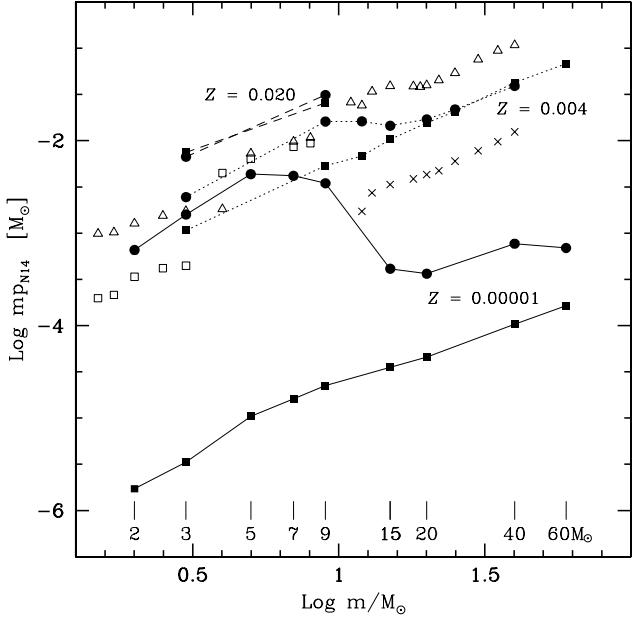}}
  \caption{Variation as a function of the initial mass of the stellar yields in $^{14}$N for different
metallicities and rotational velocities. The continuous lines refer to the models at $Z =10^{-5}$ of the present paper,
the dotted lines show the yields from the models at $Z$ = 0.004 from Maeder \& Meynet (\cite{MMVII}), the dashed lines
present the yields for two solar metallicity models (present work). The filled squares and circles indicate the cases
without and with rotation respectively. In this last case $v_{\rm ini}$ = 300 km s$^{-1}$. The crosses are for the models of 
Woosley and Weaver (\cite{WW95}) at $Z=0.1{\rm Z}_\odot$, the empty squares for the yields from van den Hoek \& Groenewegen
(\cite{ho97}) at $Z$=0.004, the empty triangles are for solar metallicity models of van den Hoek \& Groenewegen
(\cite{ho97}) up to 8 M$_\odot$ and of Woosley and Weaver (\cite{WW95})  above.
}
  \label{azotcomp}
\end{figure}

\begin{figure}[tb]
  \resizebox{\hsize}{!}{\includegraphics{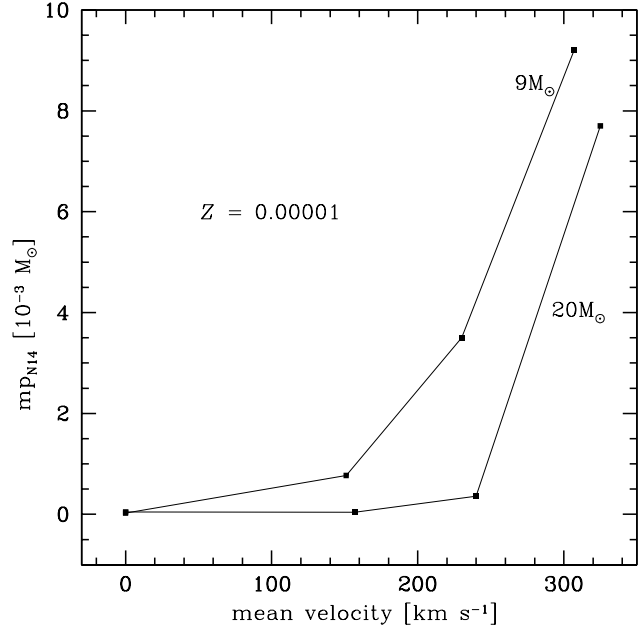}}
  \caption{Variation as a function of the mean velocity during the Main Sequence, $\overline v$, of the stellar yields in $^{14}$N
for a 9 and a 20 M$_\odot$ model at $Z = 10^{-5}$.}
  \label{vin}
\end{figure}

\section{The stellar yields}

\subsection{$M_\alpha$, $M_{\rm CO}$ and the mass of the remnants, $M_{\rm rem}$}

\begin{table}
\caption{Masses of the helium cores, of the carbon--oxygen cores and of the
remnants for
different initial mass star models with and without rotation at $Z = 10^{-5}$
and 0.004.
The masses are in solar mass and the velocities in km s$^{-1}$.
}\label{tbl-2}
\begin{center}\scriptsize
\begin{tabular}{cc|ccc|ccc}
\hline
       &            &            &              &               &            &            &                 \\
       &            &\multicolumn{3}{|c|}{$Z = 10^{-5}$}        &\multicolumn{3}{|c}{$Z = 0.004$}            \\
       &            &            &              &               &            &             &                 \\
  $M$  & $v_{\rm ini}$ & $M_\alpha$ & $M_{\rm CO}$ & $M_{\rm rem}$ & $M_\alpha$& $M_{\rm CO}$ & $M_{\rm rem}$   \\
       &            &            &              &               &            &             &                 \\
\hline
       &            &            &              &               &            &             &                 \\
 60    &     0      &  23.18     & 18.44        &  5.68         &    25.09   &   20.32     &    6.26         \\
       &   300      &  31.52     & 26.26        &  8.03         &            &             &                 \\
       &            &            &              &               &            &             &                 \\       
 40    &     0      &  14.05     & 10.50        &  3.50         &    14.61   &   10.86     &    3.59         \\
       &   300      &  15.38     & 11.54        &  3.75         &    17.87   &   14.52     &    4.49         \\
       &            &            &              &               &            &             &                 \\ 
 25    &     0      &            &              &               &     8.44   &    5.35     &    2.25         \\
       &   300      &            &              &               &     9.95   &    7.07     &    2.69         \\
       &            &            &              &               &            &             &                 \\       
 20    &     0      &   5.50     &  3.35        &  1.75         &     6.21   &    3.57     &    1.80         \\
       &   200      &   7.00     &  4.66        &  2.08         &     7.28   &    4.64     &    2.07         \\
       &   300      &   6.58     &  3.92        &  1.89         &     7.46   &    4.80     &    2.11         \\
       &   400      &   5.66     &  3.45        &  1.77         &     7.60   &    4.94     &    2.15         \\
       &            &            &              &               &            &             &                 \\       
 15    &     0      &   4.15     &  2.25        &  1.46         &     4.45   &    2.27     &    1.46         \\
       &   300      &   4.99     &  2.87        &  1.62         &     5.01   &    2.84     &    1.62         \\
       &            &            &              &               &            &             &                 \\       
 12    &     0      &            &              &               &     3.48   &    1.78     &    1.34         \\
       &   300      &            &              &               &     3.74   &    1.78     &    1.34         \\
       &            &            &              &               &            &             &                 \\       
  9    &     0      &   2.29     &  1.12        &  1.08         &     2.36   &    0.87     &    0.87         \\
       &   200      &   2.70     &  1.40        &  1.24         &            &             &                 \\
       &   300      &   2.53     &  1.28        &  1.17         &     2.85   &    1.23     &    1.14         \\
       &   400      &   2.51     &  1.31        &  1.19         &            &             &                 \\ 
       &            &            &              &               &            &             &                 \\       
  7    &     0      &   1.74     &  0.90        &  0.90         &            &             &                 \\
       &   300      &   1.07     &  1.02        &  1.02         &            &             &                 \\
       &            &            &              &               &            &             &                 \\       
  5    &     0      &   1.23     &  0.75        &  0.75         &            &             &                 \\
       &   300      &   0.88     &  0.86        &  0.86         &            &             &                 \\
       &            &            &              &               &            &             &                 \\       
  3    &     0      &   0.74     &  0.73        &  0.73         &     0.66   &     0.46    &     0.46        \\
       &   300      &   0.77     &  0.76        &  0.76         &     0.72   &     0.66    &     0.66        \\
       &            &            &              &               &            &             &                 \\       
  2    &     0      &   0.62     &  0.53        &  0.53         &            &             &                 \\
       &   300      &   0.64     &  0.56        &  0.56         &            &             &                 \\
       &            &            &              &               &            &             &                 \\ 
 \hline
\end{tabular}
\end{center}

\end{table}

\begin{table}
\caption{Stellar yields for helium, carbon, nitrogen, oxygen and the heavy
elements
from different initial mass stars 
at $Z=10^{-5}$ with and without rotation. The initial stellar masses are in
solar units. The stellar yields are in solar mass and the velocities in km
s$^{-1}$.
The contributions of the stellar winds have been accounted for and are indicated in parenthesis
when they exceed one percent of the total yield.} \label{tbl-3}
\begin{center}\scriptsize
\begin{tabular}{cc|ccccc}
\hline
             &                &                  &                 &                  &                &                \\
             &                &                         \multicolumn{5}{|c}{$Z=10^{-5}$}                               \\
             &                &                  &                 &                  &                &                \\
  $M$        & $v_{\rm ini}$  & $mp_{\rm He4}$   & $mp_{\rm C12}$  &  $mp_{\rm N14}$   & $mp_{\rm O16}$ & $mp_{\rm Z}$   \\
             &                &                  &                 &                  &                &                \\
\hline
             &                &                  &                 &                  &                &                \\
 60          &     0          &  1.1e+1          & 8.4e-1          & 1.6e-4           & 1.1e+1         & 1.3e+1         \\
             &   300          &  1.0e+1          & 6.1e-1          & 7.0e-4           & 1.6e+1         & 1.9e+1         \\
             &                & (0.2e+1)         &                 &(0.4e-4)          &                &                \\ 
             &                &                  &                 &                  &                &                \\
 40          &     0          &  6.9e+0          & 6.3e-1          & 1.0e-4           & 5.0e+0         & 7.3e+0         \\
             &   300          &  8.0e+0          & 5.8e-1          & 7.6e-4           & 5.8e+0         & 8.2e+0         \\
             &                &                  &                 &                  &                &                \\
 20          &     0          &  3.1e+0          & 2.2e-1          & 4.6e-5           & 5.6e-1         & 1.7e+0         \\
             &   200          &  2.5e+0          & 2.8e-1          & 4.2e-5           & 1.2e+0         & 2.8e+0         \\
             &   300          &  3.3e+0          & 2.9e-1          & 3.6e-4           & 9.9e-1         & 2.3e+0         \\
             &   400          &  3.5e+0          & 2.9e-1          & 7.7e-3           & 6.2e-1         & 2.0e+0         \\
             &                &                  &                 &                  &                &                \\
 15          &     0          &  2.0e+0          & 1.2e-1          & 3.5e-5           & 1.9e-1         & 8.6e-1         \\
             &   300          &  1.9e+0          & 1.9e-1          & 4.1e-4           & 3.9e-1         & 1.4e+0         \\
             &                &                  &                 &                  &                &                \\
  9          &     0          &  1.0e+0          & 1.1e-2          & 2.2e-5           & 4.0e-3         & 5.1e-2         \\
             &   200          &  1.1e+0          & 3.2e-2          & 7.7e-4           & 4.0e-3         & 2.0e-1         \\
             &   300          &  1.1e+0          & 3.6e-2          & 3.5e-3           & 4.0e-3         & 1.6e-1         \\
             &   400          &  1.1e+0          & 4.8e-2          & 9.2e-3           & 4.0e-3         & 2.0e-1         \\
             &                &                  &                 &                  &                &                \\
  7          &     0          &  6.9e-1          & 1.3e-2          & 1.6e-5           & 1.2e-3         & 1.4e-2         \\
             &   300          &  8.0e-1          & 1.8e-2          & 4.2e-3           & 5.6e-3         & 2.8e-2         \\
             &                &                  &                 &                  &                &                \\
  5          &     0          &  4.1e-1          & 6.9e-3          & 1.0e-5           & 5.6e-4         & 7.7e-3         \\
             &   300          &  4.5e-1          & 4.5e-3          & 4.3e-3           & 1.8e-3         & 1.1e-2         \\
             &                &                  &                 &                  &                &                \\
  3          &     0          &  9.0e-2          & 4.7e-4          & 3.3e-6           & 8.0e-6         & 7.0e-4         \\
             &   300          &  1.4e-1          & 2.5e-3          & 1.6e-3           & 7.0e-4         & 5.1e-3         \\
             &                &                  &                 &                  &                &                \\
  2          &     0          &  9.8e-2          & 1.5e-3          & 1.7e-6           & 1.9e-5         & 1.8e-3         \\
             &   300          &  1.2e-1          & 7.7e-3          & 6.4e-4           & 1.1e-3         & 1.0e-2         \\
             &                &                  &                 &                  &                &                \\
\hline 
\end{tabular}
\end{center}

\end{table}

\begin{table}
\caption{Same as Table \ref{tbl-3} for the metallicity $Z$=0.004.} \label{tbl-4}
\begin{center}\scriptsize
\begin{tabular}{cc|ccccc}
\hline
             &                &                  &                 &                  &                &                \\
             &                &                         \multicolumn{5}{|c}{$Z$ = 0.004}                                \\
             &                &                  &                 &                  &                &                \\
  $M$        & $v_{\rm ini}$  & $mp_{\rm He4}$   & $mp_{\rm C12}$  &  $mp_{\rm N14}$   & $mp_{\rm O16}$ & $mp_{\rm Z}$   \\
             &                &                  &                 &                  &                &                \\
\hline
             &                &                  &                 &                  &                &                \\
 60          &     0          &  9.3e+0          & 6.7e-1          & 6.8e-2           & 1.2e+1         & 1.4e+1         \\
             &                & (2.1e+0)         &                 &(1.5e-2)          &                &                \\     
             &                &                  &                 &                  &                &                \\
 40          &     0          &  6.3e+0          & 4.0e-1          & 4.3e-2           & 5.3e+0         & 7.3e+0         \\
             &                & (0.2e+0)         &                 &(0.2e-2)          &                &                \\  
             &   300          &  5.2e+0          & 4.9e-1          & 3.9e-2           & 7.3e+0         & 1.0e+1         \\
             &                & (4.1e+0)         &                 &(2.6e-2)          &                &                \\
             &                &                  &                 &                  &                &                \\
 25          &     0          &  3.7e+0          & 1.1e-1          & 2.1e-2           & 1.9e+0         & 3.1e+0         \\
             &   300          &  3.1e+0          & 2.4e-1          & 2.2e-2           & 2.6e+0         & 4.7e+0         \\
             &                & (0.1e+0)         &                 &(0.2e-2)          &                &                \\  
             &                &                  &                 &                  &                &                \\
 20          &     0          &  3.0e+0          & 8.6e-2          & 1.6e-2           & 8.5e-1         & 1.8e+0         \\
             &   200          &  2.7e+0          & 1.5e-1          & 1.7e-2           & 1.3e+0         & 2.7e+0         \\
             &   300          &  2.5e+0          & 2.1e-1          & 1.7e-2           & 1.4e+0         & 2.9e+0         \\
             &                &                  &                 &(0.1e-2)          &                &                \\ 
             &   400          &  2.5e+0          & 2.8e-1          & 1.7e-2           & 1.4e+0         & 3.2e+0         \\
             &                &                  &                 &(0.1e-2)          &                &                \\ 
             &                &                  &                 &                  &                &                \\
 15          &     0          &  2.0e+0          & 3.5e-2          & 1.0e-2           & 2.5e-1         & 8.0e-1         \\
             &   300          &  2.0e+0          & 1.9e-1          & 1.4e-2           & 3.9e-1         & 1.4e+0         \\
             &                &                  &                 &                  &                &                \\
 12          &     0          &  1.4e+0          & 8.3e-2          & 6.8e-3           & 2.7e-2         & 4.9e-1         \\
             &   300          &  1.8e+0          & 9.3e-2          & 1.6e-2           & 7.0e-2         & 6.0e-1         \\
             &                &                  &                 &                  &                &                \\
  9          &     0          &  1.2e+0          & 1.4e-2          & 5.4e-3           & 4.0e-3         & 1.6e-2         \\
             &   300          &  1.3e+0          & 4.6e-2          & 1.6e-2           & 4.0e-3         & 1.8e-1         \\
             &                &                  &                 &                  &                &                \\
  3          &     0          &  1.6e-1          & 8.8e-3          & 1.1e-3           & 6.5e-4         & 1.1e-2         \\
             &   300          &  1.3e-1          & 7.6e-4          & 2.4e-3           &-2.7e-4         & 3.6e-3         \\
             &                &                  &                 &                  &                &                \\
\hline
\end{tabular}
\end{center}

\end{table}

The computation of the stellar yields, {\it i.e.} the 
quantities of an element
newly produced by a star, necessitates, as a first step, 
the estimate of the
masses $M_\alpha$ of the helium cores, $M_{\rm CO}$ of the carbon--oxygen 
cores,  and of the masses $M_{\rm rem}$ of the remnants.
 These quantities are used for two purposes:
1) To obtain the oxygen yields using the relation between
 $M_\alpha$ and the oxygen 
yields by Arnett (\cite{ar91}). Our models only give an upper limit of the oxygen yields, 
since they are stopped at the end of
the carbon or helium burning phase, {\it i.e.} at phases where oxygen has not yet 
been depleted in the inner regions. 2) To obtain the mass of the remnants using 
a relation between $M_{\rm CO}$ and $M_{\rm rem}$. The knowledge of this quantity allows us to calculate the
mass  of the elements expelled by the supernova. In this work we use
the same $M_{\rm rem}$ vs $M_{\rm CO}$ relation as Maeder (\cite{ma92}).
For the intermediate mass stars, we have taken as remnant masses, the mass $M_{\rm
CO}$ of the CO core.

Both relations, $M_\alpha$ versus the oxygen 
yields and $M_{\rm CO}$ versus $M_{\rm rem}$
are  deduced from non--rotating models. Ideally one should have
used relations obtained from rotating models and more precisely
from rotating models treating the effects of rotation as we did.
If such computations would have been available, would the stellar yields
be the same ? For what concern helium, nitrogen and carbon (although to less extent) 
the answer is yes. Indeed the parts of the stars which contribute the most to the yields in helium, nitrogen
and carbon are in too far out portions of the star for being
affected by the $M_{\rm CO}$ versus $M_{\rm rem}$ relation.
For the oxygen and heavy elements yields the situation is more complicated.
For the moment, we can say that, unless
strong and very rapid mixing episodes take place after the end of the carbon burning
phase, it is likely that the yields in oxygen and heavy elements obtained here
are good estimates of the yields one would have obtained from 
rotating presupernovae models.

In Table~\ref{tbl-2}, $M_\alpha$, $M_{\rm CO}$
and $M_{\rm rem}$, are indicated for different initial mass
stars
with various metallicities and initial rotation velocities. For initial
masses inferior or equal to 7 M$_\odot$, the core masses are
estimated after the end of the He--burning phase, during the TP--AGB phase. For
higher initial mass models, $M_\alpha$ and $M_{\rm CO}$
are evaluated at the end of the C--burning phase at $Z$ = 10$^{-5}$, except
for the 9 M$_\odot$ models with $v_{\rm ini}$ = 0, 200 and 400 km s$^{-1}$
for which the core masses are estimated at the beginning of the C--burning phase.
For the models at $Z$ = 0.004 (Maeder \& Meynet \cite{MMVII}), the core masses 
are estimated at the end of the He--burning phase.
We define  $M_\alpha$ as the mass interior to the shell
where the mass fraction of helium becomes superior to 0.75. For the rotating
40 and 60 M$_\odot$ models at $Z = 10^{-5}$,
the  diffusion of helium outside the
He--core is so great that the above definition
yields an unrealistic high value for the helium core. For these models, we
choose to fix the outer border of the He--core at the position
where the hydrogen mass fraction goes to zero. Let us note that for the other
initial masses 
this alternative definition of the He--core does not change the results
presented in Table~\ref{tbl-2}.
The mass of the carbon--oxygen core $M_{\rm CO}$ is the mass interior to the
shell
where the sum of the mass fractions of $^{12}$C and $^{16}$O is superior to
0.75. 

From Table~\ref{tbl-2}, we note, that rotation in general increases
the masses of the helium and CO cores for the massive stars.
The reason is that for  higher  rotation, 
the intermediate convective zone, 
associated to the H--burning shell, disappears more quickly.
Since the H--burning shell is not replenished in hydrogen, it can
more quickly migrate outwards 
and thus produces the
general  increase of the He--core mass. 
However, for the models at $Z = 10^{-5}$ we notice a saturation
effect in the increase of  M$_\alpha$  for higher rotation 
and even a decrease for a very high rotation (see the 9 and 20 M$_\odot$ models
in Table~\ref{tbl-2}). 
This behaviour results from the following opposite effect:
when rotation increases, the diffusion becomes
so efficient that  large amounts of hydrogen are
brought from the radiative envelope into the H--burning shell. 
This slows down its outward progression 
and thus does not produce  the growth 
of the He--core mass in the He--burning phase.
 As to $M_{\rm CO}$, we see that
the variations of  $M_{\rm CO}$ with $v_{\rm ini}$ follows
 those of $M_\alpha$. 

In conclusion, we find that fast rotation in general 
increases $M_{\rm CO}$ and thus will also increase the yields 
in $\alpha$--elements. This is true at very low $Z$, but not
necessarily at solar metallicity, because there fast rotating
massive stars will experience high mass loss.

For the intermediate mass stars, the situation is more complicated, since in addition to the
effects just mentioned above, 
the mass of the helium core also results from the inward progression in mass of the outer convective zone 
during the AGB phase. The deeper  this inward extension, 
the smaller is $M_\alpha$.

\subsection{The stellar yields}

\begin{figure*}[tb]
  \resizebox{\hsize}{!}{\includegraphics[angle=-90]{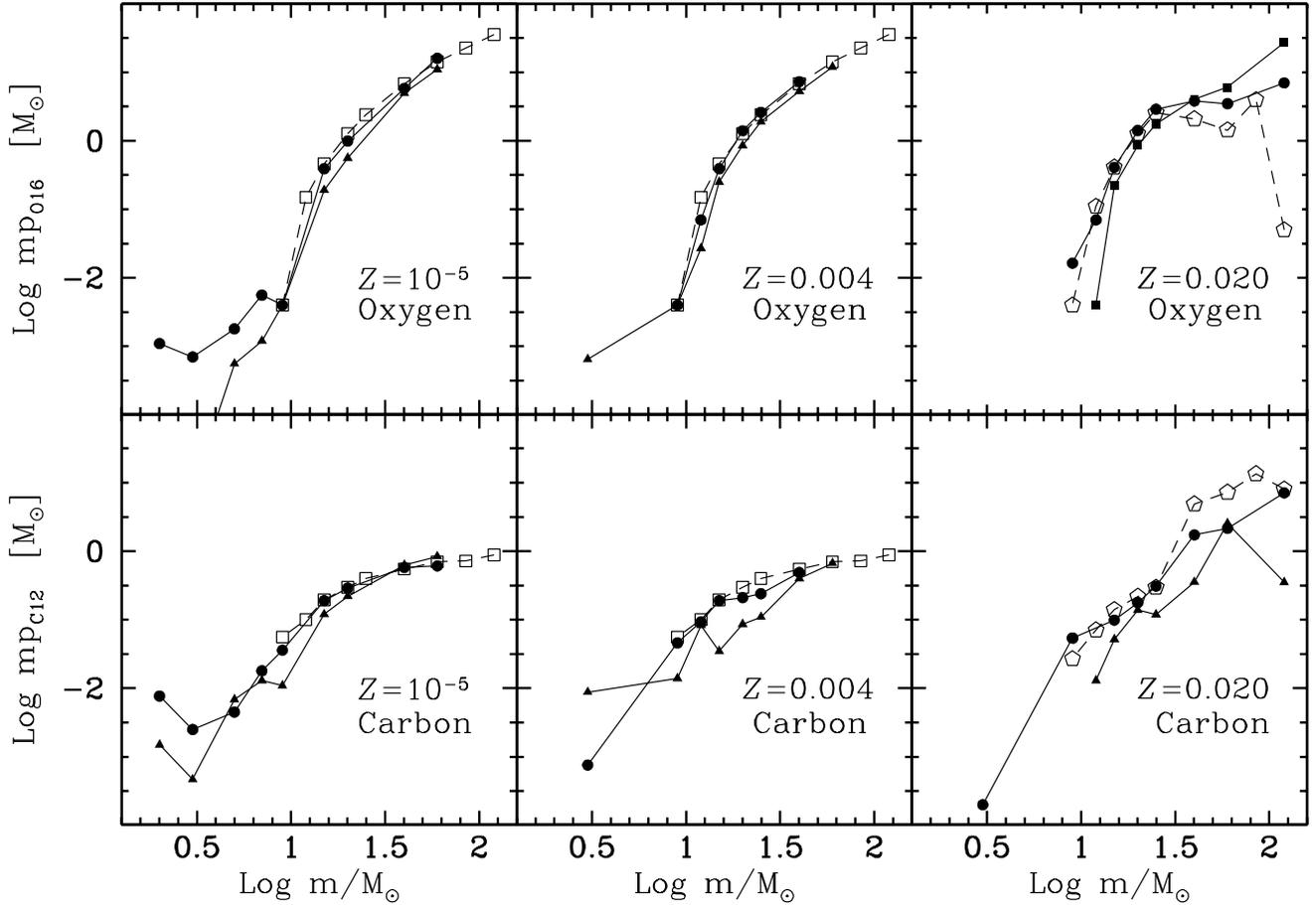}}
  \caption{Variation of the stellar yields in  oxygen and carbon as a function of the initial mass, at three different
metallicities. The black triangles are for the present non--rotating models, the black circles are for
the present rotating models. The empty squares and  pentagons are for the stellar yields of Maeder (\cite{ma92}) at the metallicity
$Z$ = 0.001 and 0.020 respectively.}
  \label{blockyield}
\end{figure*}

\begin{table*}
\caption{Integrated yields in helium, carbon, nitrogen, oxygen and heavy elements (see text).} \label{tbl-5}
\begin{center}\scriptsize
\begin{tabular}{cccccccccccc}
\hline
          &      &   &    &      &         &        &        &        &        &        &         \\
    $Z$   & 12+log(${{\rm O} \over {\rm H}}$) &  $m_{\rm d}$     & $m_{\rm u}$   &  $v_{\rm ini}$ & $P_{\rm He}$ & $P_{\rm C}$ 
& $P_{\rm N}$ & $P_{\rm O}$ &  $P_{\rm Z}$  & 
 Log ${P_{\rm C} \over P_{\rm O}}$ & Log ${P_{\rm N} \over P_{\rm O}}$  \\
          &      &   &    &      &         &        &        &        &        &        &         \\
\hline
          &      &   &    &      &         &        &        &        &        &        &         \\
10$^{-5}$ & 5.74 &20 &120 &   0  &  4.9E-3 & 3.8E-4 & 7.4E-8 & 4.1E-3 & 5.1E-3 & -1.034 & -4.743  \\        
          &      & 8 &120 &   0  &  7.8E-3 & 4.9E-4 & 1.3E-7 & 4.2E-3 & 5.8E-3 & -0.936 & -4.515  \\
          &      & 2 &120 &   0  &  1.1E-2 & 5.3E-4 & 2.1E-7 & 4.2E-3 & 5.9E-3 & -0.900 & -4.303  \\         
          &      & 2 & 60 &   0  &  9.2E-3 & 4.2E-4 & 1.8E-7 & 1.9E-3 & 3.7E-3 & -0.659 & -4.013  \\         
          &      &   &    &      &         &        &        &        &        &        &         \\
0.004$^{a}$& 8.35&20 &120 &   0  &  4.3E-3 & 2.5E-4 & 2.9E-5 & 4.7E-3 & 5.6E-3 & -1.269 & -2.205  \\        
          &      & 8 &120 &   0  &  7.3E-3 & 3.3E-4 & 4.4E-5 & 4.9E-3 & 6.4E-3 & -1.175 & -2.050  \\        
          &      & 2 &120 &   0  &  1.1E-2 & 4.6E-4 & 6.5E-5 & 4.9E-3 & 6.6E-3 & -1.024 & -1.878  \\
          &      & 2 & 60 &   0  &  9.5E-3 & 3.6E-4 & 5.5E-5 & 2.4E-3 & 4.0E-3 & -0.824 & -1.638  \\
          &      &   &    &      &         &        &        &        &        &        &         \\  
0.020$^{b}$& 8.93& 2 &120 &   0  &  8.9E-3 & 4.4E-4 & 3.6E-4 & 3.3E-3 & 5.9E-3 & -0.881 & -0.963  \\ 
          &      &   &    &      &         &        &        &        &        &        &         \\
10$^{-5}$ & 5.74 &20 &120 & 300  &  4.8E-3 & 3.3E-4 & 4.1E-7 & 6.4E-3 & 7.7E-3 & -1.286 & -4.200  \\        
          &      & 8 &120 & 300  &  7.7E-3 & 5.3E-4 & 3.8E-6 & 6.8E-3 & 9.0E-3 & -1.106 & -3.245  \\
          &      & 2 &120 & 300  &  1.0E-2 & 6.2E-4 & 1.1E-4 & 7.4E-3 & 9.5E-3 & -1.080 & -1.833  \\         
          &      & 2 & 60 & 300  &  1.0E-2 & 5.4E-4 & 3.4E-5 & 2.7E-3 & 5.0E-3 & -0.698 & -1.898  \\         
          &      &   &    &      &         &        &        &        &        &        &         \\
0.004$^{a}$& 8.35&20 &120 & 300  &  3.7E-3 & 3.8E-4 & 2.8E-5 & 7.0E-3 & 7.8E-3 & -1.265 & -2.396  \\        
          &      & 8 &120 & 300  &  6.9E-3 & 5.7E-4 & 5.8E-5 & 7.4E-3 & 9.3E-3 & -1.112 & -2.109  \\        
          &      & 2 &120 & 300  &  1.0E-2 & 6.2E-4 & 1.1E-4 & 7.4E-3 & 9.5E-3 & -1.080 & -1.833  \\
          &      & 2 & 60 & 300  &  8.8E-3 & 4.6E-4 & 9.9E-5 & 3.4E-3 & 6.0E-3 & -0.871 & -1.536  \\
          &      &   &    &      &         &        &        &        &        &        &         \\  
0.020$^{b}$& 8.93& 2 &120 & 300  &  1.1E-2 & 1.2E-3 & 3.9E-4 & 2.7E-3 & 6.5E-3 & -0.370 & -0.844  \\        
          &      &   &    &      &         &        &        &        &        &        &         \\
\hline
          &      &   &    &      &         &        &        &        &        &        &         \\
\multicolumn{12}{l}{$^{a}$ Models of Maeder \& Meynet \cite{MMVII}.} \\
\multicolumn{12}{l}{$^{b}$ Models of Meynet \& Maeder \cite{MMV}.} \\

\end{tabular}
\end{center}

\end{table*}

In Tables~\ref{tbl-3} and \ref{tbl-4}, the stellar yields for 
helium, carbon, nitrogen, oxygen and for the heavy elements
are given for different initial stellar  masses with various initial metallicities
and rotational velocities. 
Except for oxygen in massive stars, our models
have reached a sufficiently advanced evolutionary stage for the above yields to be directly
deduced from our models. The quantity of an element $x$, newly produced by a star of initial mass $m$,
{\it i.e.} the stellar yields in $x$, is given by

$$mp_{x}=\int_{M_{\rm rem}}^{M_{\rm fin}} [X_{x}(m_r)-X^0_{
x}] dm_r,$$

\noindent where $M_{\rm fin}$ is the mass of the star at the end of its evolution, $X_{x}(m_r)$
is the mass fraction
of element $x$ at the langrangian mass coordinate $m_r$ inside the star and
$X^0_{x}$ is the initial
abundance of element $x$. 
We consider that the SN ejecta in carbon consist of the C--distribution in the star
as it is at the end of the C--burning phase, but we count only the layers which are external
to the mass, where carbon is not depleted by the further nuclear burning stages
(see Maeder \cite{ma92}).

We also accounted for the effects of the stellar winds on the yields as is done in Maeder (\cite{ma92}).
In Tables~\ref{tbl-3} and \ref{tbl-4}, we have indicated in parenthesis the contribution
of the winds when it exceeds one percent of the total stellar yields.
For the metallicities considered here, the effects of the winds are small and only
modify the yields in helium and nitrogen.

In Fig.~\ref{azotcomp}, the stellar yields
 in $^{14}$N are plotted as a function of the initial mass
for different metallicities and rotation velocities. 
Our non--rotating models (filled squares along the continuous lines)
show  very small  yields at $Z$ = $10^{-5}$, as expected from a pure 
secondary origin of $^{14}$N.
When the effects of rotation are included (filled circles along the continuous lines), the yields for the intermediate mass star 
models at $Z =10^{-5}$ become of the same order of magnitude as the yields for the corresponding models at $Z$ = 0.004. 
In this mass range, the yields present thus a very weak
 metallicity dependence. 
 At solar metallicity, the yields given by the rotating and non--rotating models are identical, this well illustrates the smaller
effects of rotation on the nitrogen yields for the higher
metallicities. 

Figure~\ref{vin} 
shows that the $^{14}$N stellar yields at $Z=10^{-5}$ strongly depend on the rotational velocity. 
Starting from the point at $\overline v$ = 230 km s$^{-1}$ for the 9 M$_\odot$, 
an increase of 20\% of the mean velocity on the MS, already suffices to double the yield in nitrogen. 
One notes also the strong increase obtained for the 20 M$_\odot$ when $\overline v$ passes from 240 to 325 km s$^{-1}$.
This means that if massive stars have sufficiently high initial velocities at low $Z$, they might also play a role in the 
production of primary nitrogen (see below).

Fig.~\ref{blockyield} shows the yields in carbon and oxygen for 
$Z= 10^{-5}$, 0.004 and 0.020 for models with and without rotation. 
 We notice that 
the yields of carbon and  oxygen (and other heavy elements) are much 
less affected by rotation than the yields in nitrogen.
In general, the yields in carbon, oxygen (and  heavy elements)
are increased by rotation (cf Heger \cite{he98}), this is  a result of the generally
bigger value of M$_{\mathrm{CO}}$ when rotation is included,
 as shown above.  At solar metallicity, rotation decreases
the yield in oxygen because of the higher mass loss, which also
produces an increase in the amount of  carbon ejected. 

How the present stellar yields compare with the yields from other authors ? The situation for nitrogen can be seen
on Fig.~\ref{azotcomp}, where yields of  van den Hoek \& Groenewegen
(\cite{ho97}, see their tables 9 and 17) and of Woosley and Weaver (\cite{WW95}, see their models S and P) are plotted. 
The yields at $Z=0.004$ of
van den Hoek \& Groenewegen (\cite{ho97}) for the masses between 5 and 8 M$_\odot$ are higher by about an order of magnitude
than the yields of their lower initial mass models. They are nearly at the same level as the yields obtained from the solar metallicity models. 
This huge enhancement of their
$^{14}$N stellar yields
in this mass range is a consequence of accounting, in a parameterized way, for 
the effects of the third dredge--up and of the hot bottom burning,
 which enable the
production of primary nitrogen. 
The values of the parameters (minimum core mass for the third dredge--up, third dredge--up efficiency,
scaling law for mass loss on the AGB, core mass at which the hot bottom burning is assumed to operate)
have been chosen in order to reproduce various observational constraints, as {\it e.g.} the luminosity
function of carbon stars or the abundances observed in planetary nebulae (see van den Hoek \& Groenewegen \cite{ho97}).
Interestingly, our rotating models are well in the range of these
parameterized yields. Thus rotation, not only naturally leads to the production
of primary nitrogen, but also predicts yields in primary nitrogen
at a level compatible with those deduced from previous parameterized studies.

One also notes  that 
our yields at $Z$=0.004 (both from rotating and 
non--rotating models) are inbetween the yields
at $Z$=0.004 and $Z$=0.020 of van den Hoek \& Groenewegen
(\cite{ho97}) and between those of Woosley and Weaver (\cite{WW95}) at $Z \sim 0.002$ and 0.020. 
This indicates that at higher metallicity, 
the present yields in nitrogen  seem to be in  agreement with
the yields of other authors. Similar conclusion are reached
 when comparisons are made between our yields in
carbon and oxygen with those of these authors.
Our carbon and oxygen
yields also compare well with those obatined by 
Maeder (\cite{ma92}) (see Fig.~\ref{blockyield}).

\subsection{Net yields and comparison with the observations}

\begin{figure}[tb]
  \resizebox{\hsize}{!}{\includegraphics{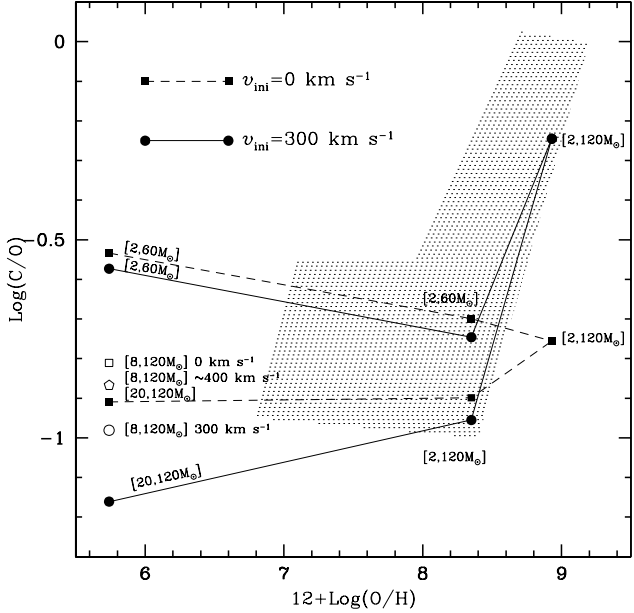}}
  \caption{Simplified model for the galactic evolution of
the C/O ratio as a function of the O/H ratio (in number). The dashed and 
continuous lines show the results deduced from the non--rotating and the rotating models respectively
(see text). The range of the initial masses used for computing the integrated yields are indicated.
The empty symbols show the results when only stars more massive than 8 M$_\odot$ are considered.
The initial velocity is indicated.
The shaded area shows the region where most 
of the observations of extragalactic HII regions and stars
are located (see e.g. Gustafsson et al \cite{gu99}; Henry et al. \cite{Ha00}).
}
  \label{cooh}
\end{figure}

\begin{figure}[tb]
  \resizebox{\hsize}{!}{\includegraphics{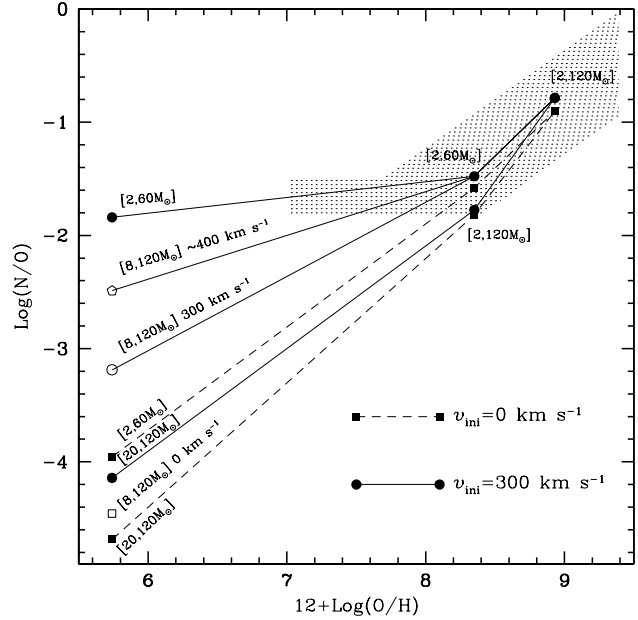}}
  \caption{Same as Fig.~\ref{cooh} for the N/O ratio.}
  \label{nooh}
\end{figure}

In order to evaluate the impact of these new yields (see Tables~\ref{tbl-3} and \ref{tbl-4}) on a galactic scale, we use them 
in a very simple model of galactic chemical evolution making use of the closed box,
instantaneous recycling approximations and supposing a constant star formation rate. We are of course fully aware  of the roughness of these approximations, but
our intention here is just to estimate the relative effects of
rotation on the chemical yields.

In the conditions of this simplified chemical evolution model, the ratio $x_i/x_j$ of the mass fractions of the elements $x_i$ 
and $x_j$ in the interstellar medium are given by 
${x_i \over x_j}= {\widetilde{y_{i}}\over \widetilde{y_{j}}}$,
where $\widetilde{y_{i}}$ and $\widetilde{y_{j}}$ are representative time--independent
approximations of the integrated yields of the elements $i$ and $j$ from a stellar generation. 
The integrated yield of
an element $x$, $P_{x}$, is defined as 
the mass fraction of all stars formed, 
which is eventually expelled under the form of the newly synthesized element
$x$:

$$P_{x}=\int_{m_{\rm d}}^{m_{\rm u}} mp_x(m) \Phi(m) dm,$$

\noindent where $\Phi(m)$ is the Initial Mass Function (IMF). Here we choose a Salpeter
IMF, normalized so that
$$\int_{0.1{\rm M}_\odot}^{120{\rm M}_\odot} \Phi(m) dm = 1,$$
The masses $m_{\rm d}$ and $m_{\rm u}$ limit the mass range of the stars having
contributed to the chemical evolution of the interstellar medium at the
epoch considered.

In Table~\ref{tbl-5}, the integrated yields are given for
various metallicities, initial rotational
velocities and values of $m_{\rm d}$ and $m_{\rm u}$. The integrated yields for the solar metallicity have been deduced from the
models of Meynet \& Maeder (\cite{MMV}). These models have been computed with a
different shear diffusion coefficient, and for a different prescription of the mass loss rate than the
present models. Therefore they do not belong to the homogeneous set of data constituted by the models at $Z$=10$^{-5}$
and 0.004. Despite these differences in the physical ingredients, their properties are well in the lines of
the results obtained at lower metallicity. This is the reason why, we have complemented the data of
Table~\ref{tbl-5} with the integrated yields obtained from these models at solar metallicity.

In Figs~\ref{cooh} and \ref{nooh}, the C/O and N/O ratios,
 obtained by simply taking the ratios of the
corresponding integrated yields, are plotted as a function of O/H.
To disentangle the still controversial role of the
intermediate mass stars  and the effects of rotation,
we have taken several values of
the upper and lower mass limits. We have considered at $Z$ = $10^{-5}$ the 
20 to 120 M$_{\odot}$ interval, that of 8 to 120 M$_{\odot}$ 
and that of  2 to 60 M$_{\odot}$. In each case, the models without
rotation and with an initial
 rotation of 300 km/s (i.e. an average of 230 km/s during the
MS phase) have been considered. In addition,
for the case of 8 to 120 M$_{\odot}$ an initial rotation of
 400  km/s has also been considered. At  $Z=10^{-5}$, 
we notice that rotation only slightly decreases the C/O ratio, the
effect is a bit larger when only massive stars are considered.
This behaviour is  due to the growth of M$_{\mathrm{CO}}$
with rotation, which favours the production of O more than 
that of C. However, we emphasize that the effect of the mass 
interval is  more important than rotation.  
When the intermediate mass stars are included,
the C/O ratio is as expected much larger. In summary, at low $Z$, the diagram
C/O vs. O/H is particularly sensitive to the mass interval.

Between $Z=0.004$ and 0.02, the main effect  influencing  
the C/O ratio is no longer 
the value of M$_{\mathrm{CO}}$ as above, but the effects of stellar
winds and their enhancement by rotation.
Rotation favours a large C/O ratio, because  rotating 
models enter at an earlier stage
into the Wolf--Rayet phase than their non--rotating counterparts.
As a consequence, in rotating models at $Z=0.02$, 
great quantities of  carbon are ejected 
by the massive stars through their stellar winds,
when they become a Wolf--Rayet star of the WC type.
This is quite in agreement with the results by
 Maeder (\cite{ma92}) who showed that
when the mass loss rates are high, most of the 
carbon is produced and ejected by massive stars
through their stellar winds. 
In non--rotating models, the new mass loss rates used
here  are much lower than those
used  in 1992 because now the mass loss rates  account for the
clumping effects in the  Wolf-Rayet stellar winds.

The diagram N/O vs O/H (Fig.~\ref{nooh}) has a different sensitivity
to the mass interval and rotation. At very low 
$Z$, we notice a very high sensitivity to rotation when 
the lowest mass limit is at 2 or 8 M$_{\odot}$; this is 
due to the production of primary nitrogen. The increase
in the N/O ratio may reach more than 2 orders of a magnitude.
Contrarily to the previous diagram, the N/O ratio is not sensitive
at all to the mass interval for models without rotation. Thus,
{\emph{the combination of the diagrams C/O vs O/H, more sensitive
to the mass interval, and of the diagram
N/O vs. O/H, more sensitive to rotation, may be particularly powerful
 to disentangle the two effects of rotation and mass interval,
 and to precise the properties of
the star populations responsible for the early chemical evolution of 
galaxies.}} 

At the present stage,
 when we compare our results to the observations we may derive the
following tentative conclusions, which could  change
if the data further improve. The C/O vs O/H diagram does not seem
favorable to  enrichments by only very massive stars in the
range 20 to 120 M$_{\odot}$; contributions from stars down 
to 8 or 2 M$_{\odot}$ may be needed, depending on the exact
 slope observed at low $Z$ in Fig.~\ref{cooh}.
As to the  N/O vs O/H diagram, no model without rotation
is able to account for the observed
plateau, moreover contributions from only stars above 
20 M$_{\odot}$ seem difficult. The observed plateau at 
$\log$ N/O = -1.7 strongly supports rotating models  including the 
large contribution from  intermediate mass stars down  to, either 
 2 M$_{\odot}$ if these stars have the same average rotation
as in Pop. I stars, or down to only 8 M$_{\odot}$ if the rotations are
faster as suggested by Maeder et al. (\cite{mgm}).
In this respect, it would be sufficient that the average 
rotational velocities during the MS phase are larger by about 80 km/s.

We must temperate these conclusions by the 
following remark (cf. also Meynet \& Maeder \cite{MMlettreN}),
related to a current problem in the chemical evolution of 
galaxies. Nitrogen is ejected mainly by AGB stars with 
ejection velocities 
of a few 100 km s$^{-1}$, while oxygen is ejected by supernovae  
at much higher velocities of $10^4$  km s$^{-1}$ or more. Thus, 
a fraction of the oxygen produced may escape from the parent galaxy, 
leading to a higher N/O ratio than in the simple estimate made here.

\section{Conclusions}

We have investigated with some details the physics of
very low $Z$ models with  $Z$ = 10$^{-5}$ . Even if we have 
little chance before long  to observe such star populations, these 
models are most relevant for the early chemical evolution
of galaxies. The models
 enable us to calculate the yields in CNO, He and
heavy elements necessary to study the early evolution
of galaxies. The diagrams showing log N/O vs log O/H
and log C/O vs O/H are particularly powerful to infer properties
of the early star generations in galaxies.

This work shows the large  differences brought by 
the present models with rotation. 
In future we shall
calculate the detailed yields of
 the various heavy elements.

\end{document}